%% file: main.tex
\documentclass[journal]{IEEEtran}
\usepackage{amsthm}
\usepackage{cite}
\usepackage{amsmath}
\usepackage{nccmath}
\usepackage{amssymb}
\usepackage{amsfonts}
\usepackage{graphicx}
\usepackage{fancyhdr}
\usepackage{svg}
\usepackage{bbm}
\usepackage{textcomp}
\usepackage{xcolor}
\usepackage{algorithmicx,algpseudocode}
\usepackage[linesnumbered,ruled]{algorithm2e}
\usepackage{booktabs}
\usepackage{bbold}
\usepackage{kotex}
\usepackage{mwe}
\usepackage{multicol}
\usepackage{multirow}
\usepackage[normalem]{ulem}
\usepackage[bottom]{footmisc}
\usepackage{mwe}
\usepackage{color}
\usepackage{booktabs,multirow}
\usepackage{tabularx}
\usepackage{tikz}
\usepackage{kotex}
\usetikzlibrary{patterns,arrows}
\usepackage{physics}
\usepackage{verbatim}
\usepackage{subfig}
\usepackage{circledsteps}
\usepackage{microtype}
\usepackage[export]{adjustbox}

\theoremstyle{definition}

\definecolor{color1}{RGB}{228,26,28}
\definecolor{color2}{RGB}{55,126,184}
\definecolor{color3}{RGB}{77,175,74}
\definecolor{color4}{RGB}{152,78,163}
\definecolor{color5}{RGB}{255,127,0}
\definecolor{color6}{RGB}{241,241,241}
\definecolor{color7}{RGB}{156,156,156}
\definecolor{color8}{RGB}{96,96,96}
\definecolor{color0}{RGB}{162, 20, 47}

\newcommand{\tpurp}[1]{{\color{black}{#1}}}

\newcommand{\BfPara}[1]{{\noindent\bf#1.}\xspace}

\newcommand*\circled[1]{\Circled[inner color=white, fill color= color0, outer color=color0]{\footnotesize{#1}}}

\renewcommand{\baselinestretch}{.98}

\begin{document}

\title{\fontsize{23}{28}\selectfont Handover Protocol Learning for LEO Satellite Networks: Access Delay and Collision Minimization}

\author{
    Ju-Hyung Lee,~\IEEEmembership{Member, IEEE}, 
    Chanyoung Park, 
    Soohyun Park, 
    and 
    Andreas F. Molisch,~\IEEEmembership{Fellow, IEEE}
    \thanks{J.-H. Lee and A. F. Molisch are with Ming Hsieh Department of Electrical and Computer Engineering, University of Southern California, Los Angeles, USA (Emails: \{juhyung.lee, molisch\}@usc.edu).}
    \thanks{C. Park and S. Park are with the School of Electrical Engineering, Korea University, Seoul 02841, Korea (Emails: \{cosdeneb,soohyun828\}@korea.ac.kr).}
    \thanks{S. Park is a corresponding author of this paper.}
}
\maketitle

\input{Section/Abstract.tex}

\section{Introduction} \label{sec:introduction}

\input{Section/SEC1.tex}



\section{Related Works} \label{sec:relatedworks}
\input{Section/SEC1-B.tex}

\section{Handover for LEO SAT-based Non-Terrestrial Networks} \label{sec:background}

\input{Section/SEC2.tex}

\input{Section/SEC3.tex}

\section{DRL-based Handover Protocol for LEO Satellite Networks} \label{sec:DHO}
\input{Section/SEC4.tex}

\section{Numerical Results} \label{sec:simulation}
\input{Section/SEC5.tex}

\section{Conclusion} \label{sec:conclusion}
\input{Section/Conclusion.tex}

\begin{appendices}
\section{Selection of State Information}  \label{sec:appendix}
\begin{figure}[!h]
\centering
\subfloat[DHO-Centralized. \label{fig:importantInfo_DHO_Centralized}]{\includegraphics[width=.7\linewidth]{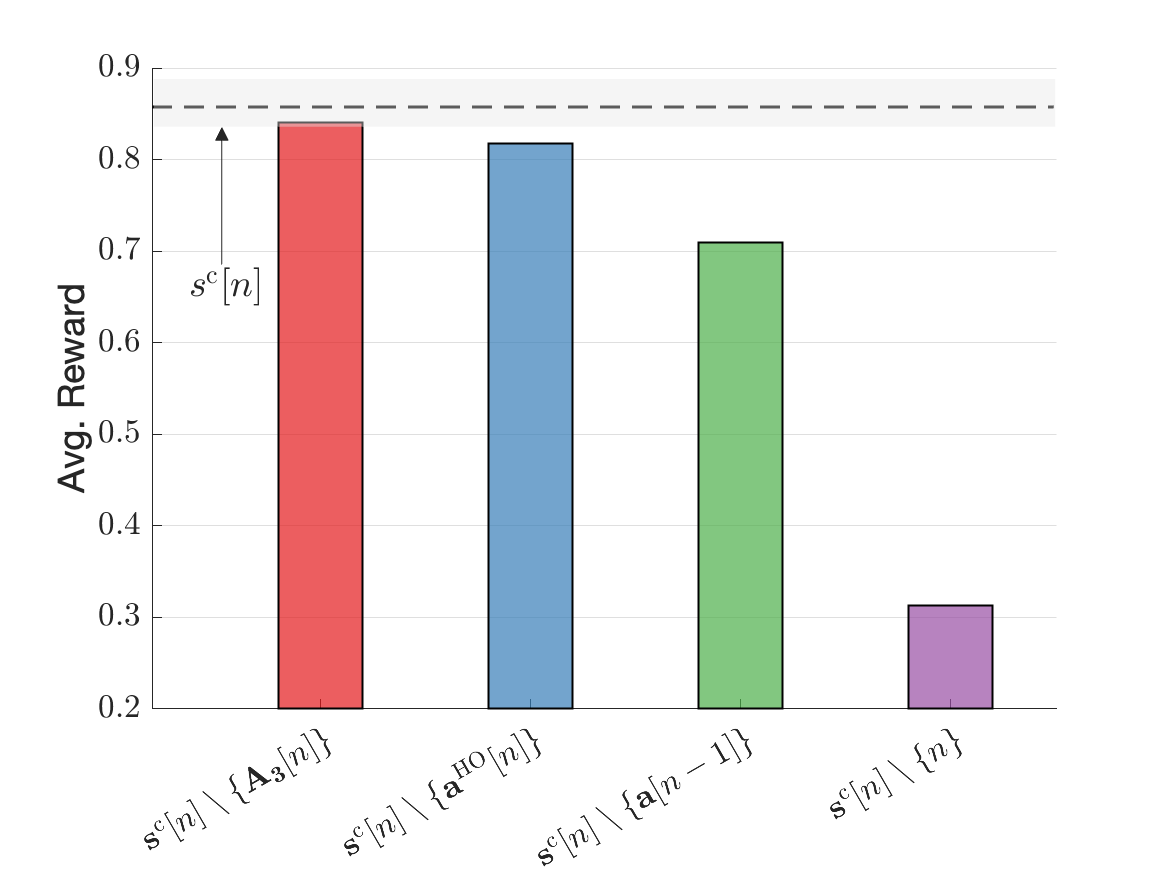}}\vspace{-1.em}\\
\subfloat[DHO. \label{fig:importantInfo_DHO}]{\includegraphics[width=.7\linewidth]{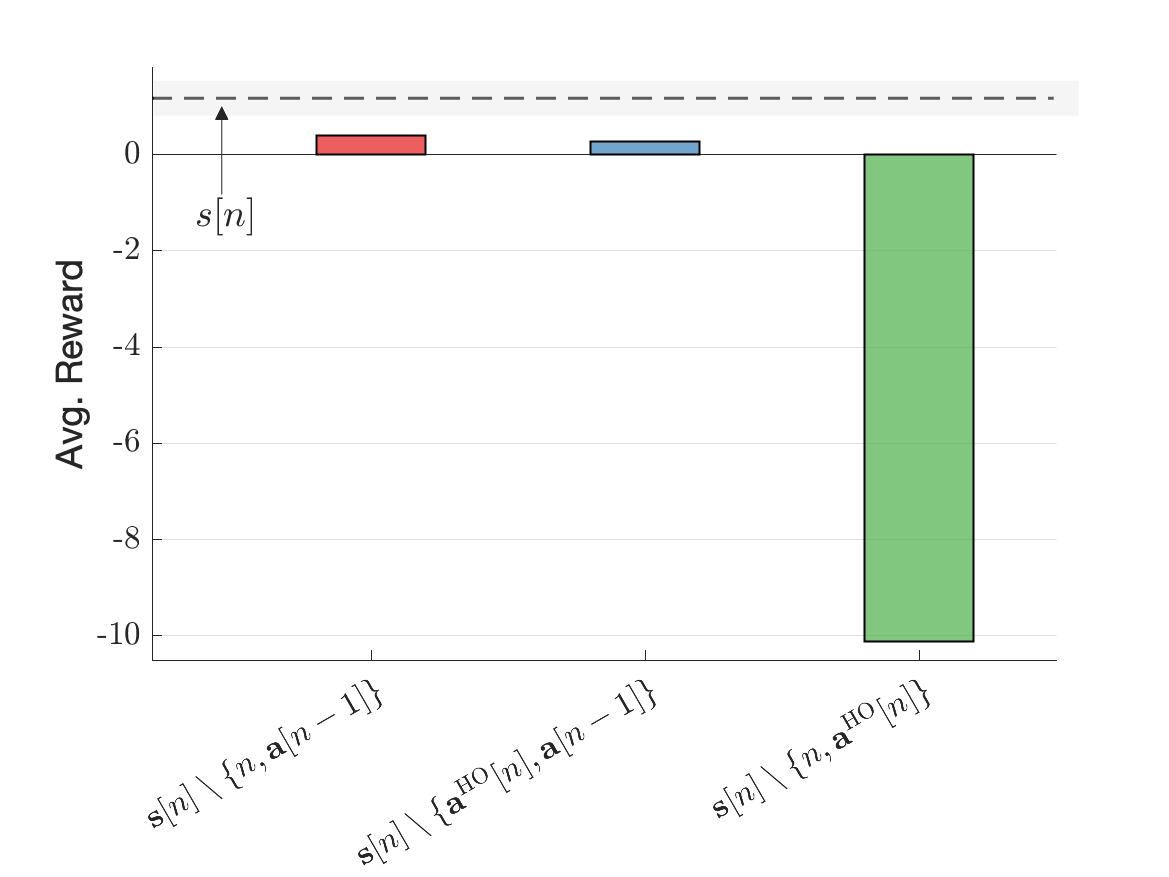}}
\caption{Impact of each information included in $\mathbf{s}[n]$.}
\label{fig:comparion_stateInfo}
\vspace{-.5em}
\end{figure}

The state space of an agent should contain sufficient information for decision-making while minimizing additional data collection overhead from the environment to promote DRL training convergence. To identify the important information to include in the state space, we conducted an extensive study to evaluate the impact of various locally observable information on the performance of the proposed DHO protocol. Our findings are presented in an ablation study in Fig. \ref{fig:comparion_stateInfo}, which illustrates the contribution of each piece of information to the overall system.

We first study the centralized case, referred to as DHO-Centralized, as shown in Fig. \ref{fig:importantInfo_DHO_Centralized}. In this case, the state information includes both locally observable and non-observable (centralized) information (\textit{e.g.}, A3 event $\mathbf{A_3}[n]$), which is defined as:
\begin{align}
\mathbf{s}^{\mathrm{c}}[n] &= \lbrace \mathbf{s}[n], \mathbf{A_3}[n] \rbrace. \label{State_C}
\end{align}
Here, the performance of DHO-Centralized serves as an upper-bound result of our proposed DHO scheme.
As shown in the figure, the DHO protocol achieves near-optimal performance while minimizing complexity.
Note that $\mathbf{s}^{\mathrm{c}}[n] \setminus \mathbf{A_3}[n]$ is equivalent to $\mathbf{s}[n]$ and that DHO utilizes only locally observable information.

Secondly, Fig. \ref{fig:importantInfo_DHO} demonstrates that the DHO protocol primarily depends on two locally observable pieces of information: 1) the time index and 2) the accessed UEs. These two pieces of information have a significant impact on the training of DHO. Interestingly, that information is locally observable which highlights the potential for a distributed multi-agent DRL approach in future work. 

\section{Selection of DRL algorithm}  \label{sec:appendixB}



\begin{figure*}[!ht]
\vspace{-1.em}
\centering
\subfloat[\tpurp{DQN, PPO, A3C, and IMPALA} \label{fig:comparison_DRL1}]{\includegraphics[width=0.35\linewidth]{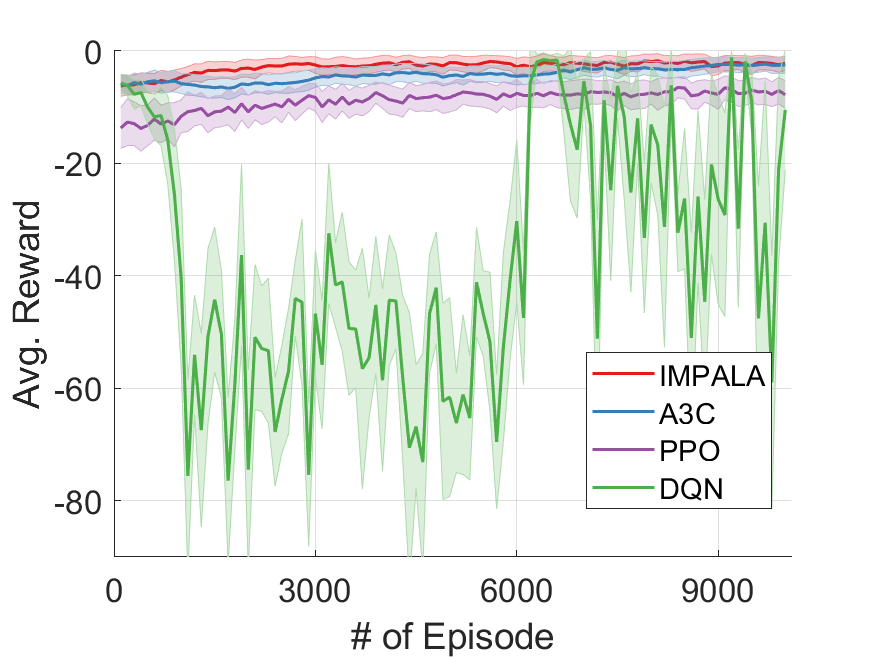}}
\subfloat[\tpurp{PPO, A3C, and IMPALA} \label{fig:comparison_DRL2}]{\includegraphics[width=0.35\linewidth]{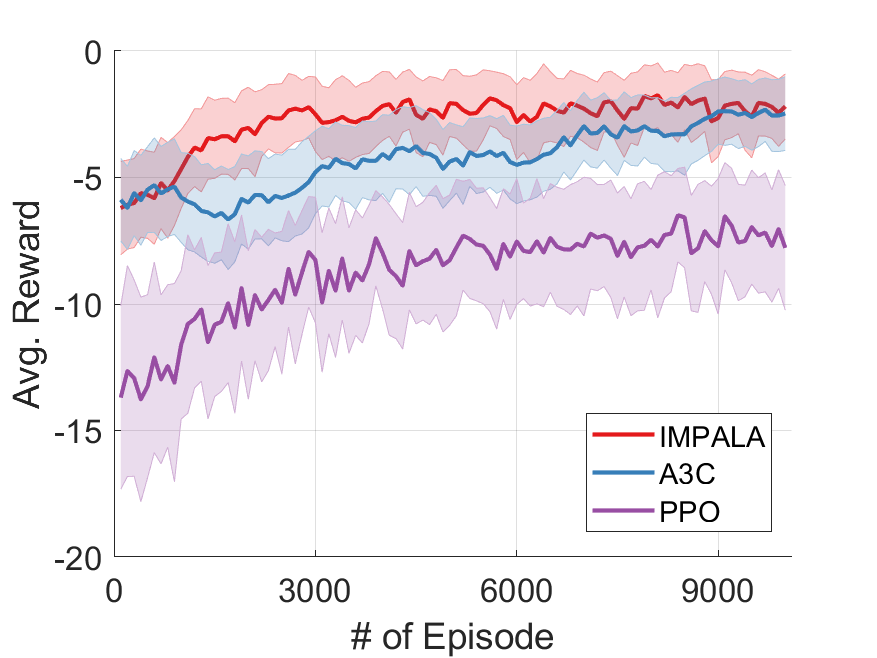}}
\caption{\tpurp{Comparison of DRL algorithms ($J=10$, $R_k = 0.3 J, \ \forall k$, and $P=5 J$). The shaded region represents the variance, while the line represents the average of the cumulative reward for an episode.}}
\label{fig:comparion_DRL}
\vspace{-1.em}
\end{figure*}

\tpurp{
We empirically demonstrate the superior performance of IMPALA over DQN, PPO, and A3C in our specific environment. Fig. \ref{fig:comparion_DRL} showcases the convergence behavior of the three algorithms and provides insights into their performance characteristics.
Firstly, as shown in Fig. \ref{fig:comparison_DRL1}, we observe that DQN faces challenges in achieving convergence due to the large action space. Unlike IMPALA, A3C, and PPO, which can handle multi-discrete type actions, DQN requires alternative approaches, such as flattening the action space, to select an action from all possible actions.
Secondly, as shown in Fig. \ref{fig:comparison_DRL2}, IMPALA exhibits better performance compared to A3C and PPO in terms of stable convergence, thanks to its scalability and improved sample efficiency (see Sec. \ref{sec:algorithmDetail}). These results highlight the advantages of IMPALA over other DRL algorithms in our study.
}
\end{appendices}

\bibliographystyle{IEEEtran}
\bibliography{ref_main, ref_aimlab, ref_twc}

\end{document}

%% file: Section/Abstract.tex
\begin{abstract}
\tpurp{
This study presents a novel deep reinforcement learning (DRL)-based handover (HO) protocol, called DHO, specifically designed to address the persistent challenge of long propagation delays in low-Earth orbit (LEO) satellite networks' HO procedures. DHO skips the \texttt{Measurement Report} (MR) in the HO procedure by leveraging its predictive capabilities after being trained with a pre-determined LEO satellite orbital pattern. This simplification eliminates the propagation delay incurred during the MR phase, while still providing effective HO decisions. The proposed DHO outperforms the legacy HO protocol across diverse network conditions in terms of access delay, collision rate, and handover success rate, demonstrating the practical applicability of DHO in real-world networks. Furthermore, the study examines the trade-off between access delay and collision rate and also evaluates the training performance and convergence of DHO using various DRL algorithms.
}


\end{abstract}

\begin{IEEEkeywords}
LEO satellite network, handover, protocol learning, deep reinforcement learning, 6G.
\end{IEEEkeywords}
\IEEEpeerreviewmaketitle

%% file: Section/SEC1.tex

In the realm of beyond-5G and 6G networks, satellite constellations utilizing low-Earth orbit (LEO) and medium-Earth orbit (MEO) satellites (SATs) have emerged as a crucial solution for providing global coverage. These constellations, composed of thousands of SATs in close proximity to the Earth, enable high-speed, low-latency broadband internet access in remote and underserved areas across the globe \cite{Starlink, FCC_Kuiper}.
However, as the number of SATs in mega-constellation networks continues to rise and these networks are increasingly utilized for a diverse range of applications, there is a corresponding increase in the number of devices attempting to access the network, known as massive access. This can lead to a potential for network congestion and an increase in latency, which is a significant concern in these networks \cite{SAT_Survey4, SAT_Survey5, SAT_Survey1}.


Since massive access may negatively impact network performance, new approaches are required to handle it \cite{SAT_UCL_1}.
Conventional \emph{transparent-type} LEO SAT, which relies on centralized reservation methods, where a limited number of ground stations manage various functions such as handover (HO) and resource allocation for all user equipment (UEs) connected to the mega-constellation, may not be efficient in handling this massive access scenario. To overcome this limitation, \emph{regenerative-type} LEO SAT has emerged, which is capable of making decisions and performing functions without relying on ground stations\footnote{Transparent-type SATs function by amplifying and forwarding signals, while regenerative-type SATs possess the capability for on-board processing (OBP) and are able to perform functions, such as HO decision \cite{3GPP_NTN2}.} \cite{3GPP_NTN2}. However, applying traditional HO protocols used in terrestrial networks to regenerative-type LEO SAT networks may not be optimal as these protocols were not designed for the high dynamics of LEO SAT networks or the long propagation delay encountered in them, leading to unnecessary delay and power consumption from burdensome HO procedures.

In this paper, we propose a novel HO protocol specifically designed for the regenerative-type LEO SAT networks using the deep reinforcement learning (DRL) approach.
Our contributions can be summarized as follows:
\begin{itemize}

\item A novel HO protocol is proposed for the LEO SAT networks, called DHO.
DHO re-designs conventional HO procedures to suit LEO SAT network requirements. The protocol leverages locally observable information, tailored to the unique characteristics of LEO SAT networks.

\item \tpurp{The proposed DHO protocol minimizes access delay and collision rate while simplifying the HO process by skipping the \texttt{Measurement Report} (MR), achieving better performance even with lower power consumption.
The DHO employs the importance-weighted \textit{Actor}-\textit{Learner} architecture, IMPALA algorithm \cite{espeholt2018impala}, which ensures stable training of cases with large state and action spaces (see Sec.~\ref{sec:complexity}) compared to other DRL algorithms (see Appendix \ref{sec:appendixB}).}

\item Numerical results corroborate that the DHO protocol demonstrates its superiority under various conditions over conventional HO protocol, achieving up to $6.86$x and $4.18$x lower access delay than conventional HO and heuristic methods, respectively, (see \textbf{Table \ref{table:comparison_RB} and \ref{table:comparison_PRACH}} in Sec.~\ref{sec:simulation}). 
Also, the DHO agent behavior (trained policy) reveals the underlying mechanism for its improved performance and its adaptability in various network scenarios (see \textbf{Table \ref{table:agentPolicy}} in Sec.~\ref{sec:simulation}).

\item Furthermore, the versatility of the DHO protocol in adapting to different network scenarios is also demonstrated (see \textbf{Table \ref{table:coeffRatio_Reward}} in Sec.~\ref{sec:simulation}).

\end{itemize}

The rest of the paper is organized as follows: After discussing the related works on HO processes for LEO SAT-based non-terrestrial networks (NTN) in Sec.~\ref{sec:relatedworks}, Sec.~\ref{sec:background} provides a background and covers the challenges of HO in LEO SAT networks. Sec.~\ref{sec:DHO} presents the proposed DHO algorithm and its evaluation method. Sec.~\ref{sec:simulation} provides simulation results and performance analysis, followed by concluding remarks in Sec.~\ref{sec:conclusion}.

\textit{Notation:}
Throughout this paper, we use the normal-face font to denote scalars, and boldface font to denote vectors.
We use $\mathbb{R}^{D\times 1}$ to represent the $D$-dimensional space of real-valued vectors.
$\nabla_{\mathbf{x}} f(\mathbf{x})$ denotes the gradient vector of function $f(\mathbf{x})$. 

%% file: Section/SEC1-B.tex
Various HO strategies have been proposed to address the challenges posed by HO in LEO SAT networks. One such strategy employs a graph theory-based approach, where the relationship between LEO SAT and UE is represented as a graph and node~\cite{7470617, 8836603, 9068245,jhl4}. While this graph-based approach has the potential to find efficient HO decisions, it is only applicable to the central reservation method in transparent-type LEO SAT networks, but not for the regenerative-type LEO SAT networks, which require real-time decisions to be made through on-board processing.

To address this limitation, DRL can be used to make real-time decisions. The use of DRL in wireless communications has been extensively studied, as reviewed in~\cite{8714026}. For instance, in dynamic spectrum sensing/access in cognitive radio, DRL has been applied to determine appropriate spectrum access strategies~\cite{9557877, 8474348, 9089307}. Furthermore, the use of DRL can be extended to random access performance improvements in LEO SAT networks~\cite{jhl1,jhl2,jhl3}.
In the context of LEO SAT networks, some studies have applied DRL to improve resource allocation and capacity management, which, however, primarily focus on central decisions~\cite{8713802, P4J_JH, 9067004}.

Several studies have also investigated the application of machine learning in HO procedures \cite{9206115, 9221072, 9223658, 8959359}. 
For instance, \cite{9221072} utilizes a DRL methodology in certain HO processes, although in a centralized manner. \cite{9223658} aims to optimize both beamforming and HO jointly but only addresses the HO process at a high level without consideration for detailed HO procedures. \cite{8959359} employs a supervised learning approach to enhance the predictive capability of a specific HO procedure; however, this approach yields limited improvements and heavily relies on a data-driven approach. Although previous studies demonstrate the potential of machine learning integration in the HO process, they primarily employ methods that are either too high-level or overly specialized; besides, these studies do not sufficiently account for the dynamic environments typical of NTN.

Considering the outlined approaches, our research aims to enhance the DRL-based approach, particularly focusing on HO in LEO SAT networks. We propose a novel DRL-based HO protocol designed to enable efficient real-time decision-making, thereby broadening the scope of DRL-based solutions within LEO SAT network contexts.

%% file: Section/SEC2.tex
\begin{figure}[t!]
    \centering
    \includegraphics[width=.84\linewidth]{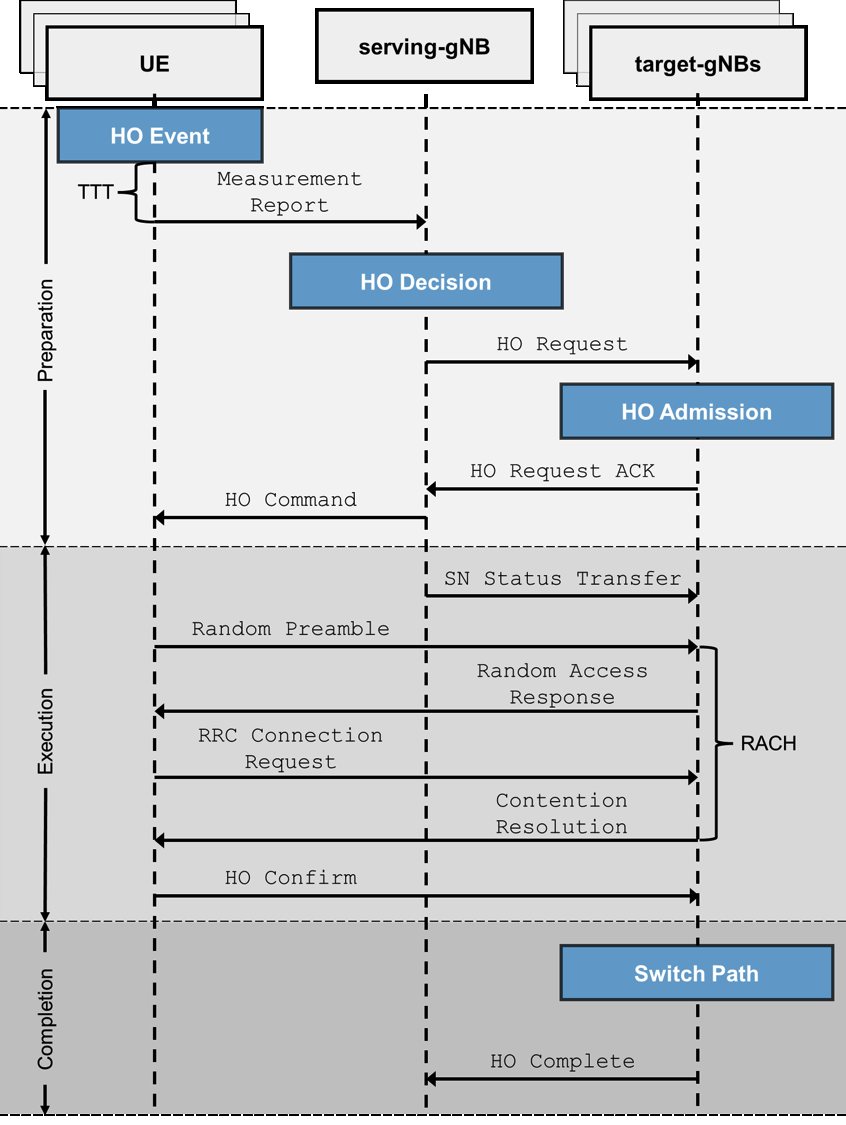}
    \caption{The sequence diagram of conventional HO procedure.}
    \label{fig:existing HO process}
    \vspace{-1.em}
\end{figure}

\begin{table*}[!h]
\centering
\caption{The entering conditions of HO event.}
\resizebox{.95\linewidth}{!}{\begin{minipage}[t]{1\textwidth}
\centering
\renewcommand{\arraystretch}{1.0}
\label{table:A1_5}
\centering
\input{Table/Table_HOevent.tex}
\centering
\end{minipage}}
\vspace{-1.em}
\end{table*}

\subsection{Overview of HO Process} \label{sec:HO_overview}
As mentioned previously, \emph{transparent-type} LEO SATs mainly operate on schedule-based HO mechanisms, such as centralized reservation methods. In these methods, a limited number of ground stations manage various functionalities, including HO and resource allocation for all UEs connected to thousands of LEO SATs.
However, in this paper, our primary focus is on another type of platform, \emph{regenerative-type} LEO SAT. This type of platform is capable of making autonomous decisions and employs a trigger-based HO process, corresponding to the conventional HO process utilized in the 3GPP New Radio (NR).

HO in 3GPP NR, the primary 5G cellular standard, consists of three main phases: preparation, execution, and completion~\cite{8959359}. This process, shown in Fig.~\ref{fig:existing HO process}, ensures the continuous transfer of data sessions or calls when a UE moves from one cell to another \cite{Wirelee_Book_AM}.



\BfPara{Preparation phase}
During the \textit{preparation phase}, the UE measures signals from the serving and target Node Bs (gNBs), deciding if a HO event is necessary based on five potential conditions outlined in Table \ref{table:A1_5}. The handover margin (HOM) and time-to-trigger (TTT) are used to evaluate these conditions. If satisfied, the UE sends a \texttt{MR} to the serving-gNB, leading to a \texttt{HO Request} to the target-gNB. If accepted, the serving-gNB sends a \texttt{HO Command} to the UE, moving to the \textit{execution phase}.

\BfPara{Execution phase}
In the \textit{execution phase}, the serving-gNB transfers the UE's data and a sequence number to the target-gNB~\cite{alexandris2016analyzing}, and the UE connects to the target-gNB through a random access channel (RACH) process~\cite{hasan2013random, 7878690}. The UE sends a \texttt{HO Confirm} message when the connection is successful.


\BfPara{Completion phase}
The \textit{completion phase} includes the network entities' request to switch the packet path from the serving-gNB to the target-gNB, releasing the resources from the serving-gNB once the \texttt{HO Complete} message is transmitted.


The HO process, particularly the preparation phase, is vulnerable to failure under conditions of poor signal quality, where issues such as transmission/reception failure of HO-related messages or radio link failure (RLF) can occur. Considering the increased vulnerability in the preparation phase in LEO SAT-based NTN, which often faces challenging channel conditions due to their ultra-long link distance, our work focuses on enhancing this initial stage of the HO process.


%% file: Table/Table_HOevent.tex
\begin{tabular}{c|c}
\toprule[1pt]
\textbf{Event} & \textbf{Entering Condition}\\ \midrule
A1 & serving-gNB's signal becomes stronger (better) than the threshold\footnote{
Event A1 is mainly used to stop the measurement of a certain cell rather than triggering the HO.
}.\\
A2 & serving-gNB's signal becomes weaker (worse) than the threshold.\\
A3 & target-gNB's signal becomes stronger (better) than serving-gNB's signal by a margin of the offset value.\\
A4 & target-gNB's signal becomes stronger (better)  than the threshold.\\
A5 & serving-gNB's signal becomes weaker than threshold 1, and target-gNB's signal becomes stronger than threshold 2.\\
\bottomrule[1pt]
\end{tabular}

%% file: Section/SEC3.tex
\subsection{Challenges in LEO SAT-based NTN}
In LEO SAT-based NTN, HO poses unique challenges compared to terrestrial cellular networks. These challenges include:
\begin{enumerate}
    \item \emph{Long propagation delay}: The signal from a UE to an SAT and back to a ground station takes a significant amount of time, leading to delays in the HO process.
    \item \emph{Large coverage area}: The coverage area of SAT is much larger than that of a terrestrial cell, which can make it more difficult to identify the best target gNB for HO.
    \item \emph{Limited resources}: The limited resources available on SAT can make it harder to support HO, especially when multiple UEs need to conduct HO at the same time.
\end{enumerate}
As a result of these challenges, HO in LEO SAT networks is considered a more challenging process than the one in terrestrial cellular networks.

\begin{figure*}[!h]
\centering
\subfloat[Propagation delay $D_{\mathrm{P}}$.]{
\includegraphics[width=0.28\linewidth]{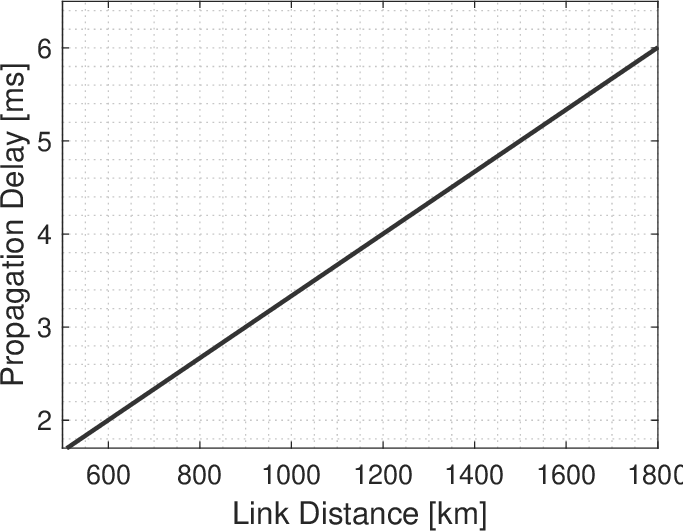}
\label{fig:Propagation_delay}
}
\subfloat[CNR (VSAT).]{
\includegraphics[width=0.28\linewidth]{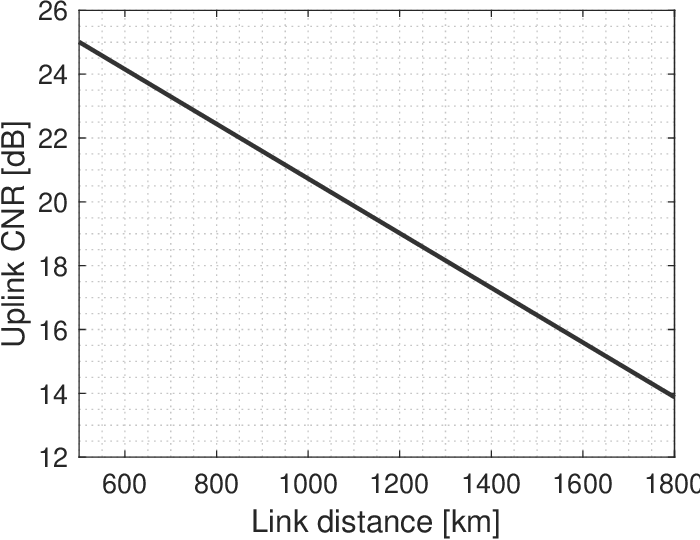}
\label{fig:CNR_VSAT}
}
\subfloat[CNR (Handheld).]{
\includegraphics[width=0.28\linewidth]{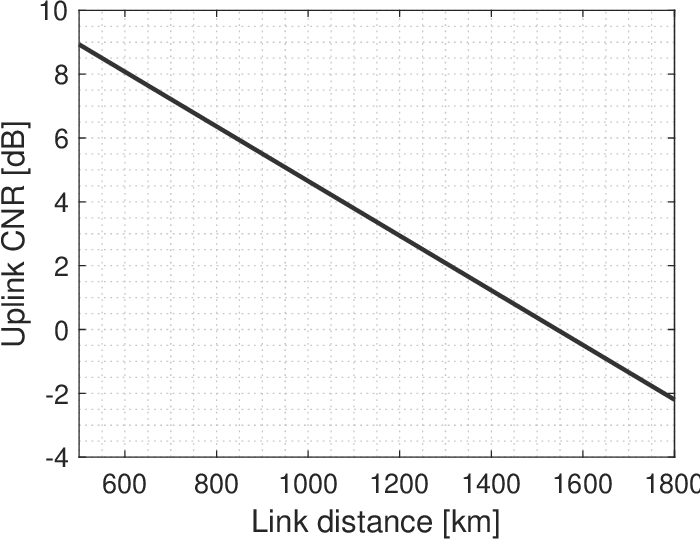}
\label{fig:CNR_Handheld}
}
\caption{Propagation delay, CNR, and BER in networks between ground UE and LEO SAT. Each result is calculated by the parameters specified by~\cite{3GPP_NTN2}, respectively.}
\label{fig:challenge}
\vspace{-1.em}
\end{figure*}

\subsubsection{Challenge: Outdated, Power-Consuming, and Unreliable \texttt{MR}}
Specific challenges in LEO SAT networks are the outdated, power-consuming, and unreliable \texttt{MR}. 
The transmission of the signaling is hampered by a long propagation delay, which is a result of the physical distance between the ground UE and the LEO SAT providing network services at an altitude of $500\sim2000$ [km].
As the propagation speed is constant (\textit{e.g.}, $c=2.997 \times 10^{8}$), the propagation delay is proportional to the distance between the two terminals. The one-way propagation delay can be as long as $1.6\sim6$ [ms], as shown in Fig. \ref{fig:Propagation_delay}, which can result in \texttt{MR} sent from ground UE to LEO SAT becoming outdated.

\tpurp{
Additionally, the long link distance between the UE and LEO SAT necessitates high uplink (UL) signal power consumption during the \texttt{MR}.
By calculating the carrier-to-noise ratio (CNR), it is possible to see the required UL transmit power for the UE to periodically transmit signaling to the LEO SAT to report the HO condition and also to estimate its reliability.
}

For the NTN scenario, CNR can be calculated as:
\begin{align}
\begin{split}
    \mathrm{CNR\;[dB]} =& \ \mathrm{EIRP\;[dBW]} -P_{\mathrm{FS}}\;\mathrm{[dB]} -P_{\mathrm{A}}\;\mathrm{[dB]} \\
    &  -P_{\mathrm{SM}}\;\mathrm{[dB]} -P_{\mathrm{SL}}\;\mathrm{[dB]} +\; G/T\;\mathrm{\;[dB/K]} \\
    & -\;k\mathrm{\;[dBW/K/Hz]} - \;B\mathrm{\;[dBHz]},
    \label{eq:CNR}
\end{split}
\end{align}
where $\mathrm{EIRP}$ is the equivalent, isotropically radiated power (EIRP) in the transmitter (TX); $P_{\mathrm{FS}} = 20 \log_{10}(f) + 20 \log_{10}(d) + 92.45$ with being the carrier frequency $f$ in GHz and the link distance $d$ in km; $P_{\mathrm{A}}$ is the atmospheric path loss due to gases and rain fades in between groud-to-space; $P_{\mathrm{SM}}$ is a shadowing margin; $P_{\mathrm{SL}}$ is scintillation loss;  $G/T$ is is antenna-gain-to-noise-temperature in the receiver (RX); $k$ is the theoretical noise floor or the minimum sensitivity of the receiver as the Boltzmann constant; $B$ is the channel bandwidth. 
Here, for the UL communication between the LEO SAT at an altitude of 600 [km] and the ground UE, the $\mathrm{S}$-band (2 [GHz]) is considered for handheld-type UE and $\mathrm{K_a}$-band (30 [GHz]) is considered for VSAT-type UE. 
Note that the parameter values for \eqref{eq:CNR} are summarized in Table.~\ref{tab:terminal_parameter}.

\begin{table}[]
\caption{Parameters of UL transmission in LEO SAT networks (see Tables 6.1.1.1-1 and 6.1.1.1-3 of~\cite{3GPP_NTN2}).}
\label{tab:terminal_parameter}
\centering
\resizebox{.95\linewidth}{!}{\begin{minipage}[t]{0.5\textwidth}
\centering

\input{Table/Table_NTNparameter_3GPP.tex}

\end{minipage}}
\vspace{-1.em}
\end{table}

Fig. \ref{fig:challenge} highlights the challenges faced in the traditional HO process, which can result in significant power consumption for the UE in the uplink transmission, as well as low reliability, due to the procedure of sending the \texttt{MR}.

\subsubsection{Challenge: High Correlation and Density of Ground UEs}
In LEO SAT networks, handling massive HO requests for UEs in densely populated areas poses another challenge, as there is a high correlation among UEs in such areas, as shown in Fig. \ref{fig:networkscenario}. 
For instance, in scenarios where ground UEs are situated in a street canyon, they may encounter simultaneous loss and gain of line-of-sight due to the overhead passage of the SAT. As a consequence, the probability of collision increases when multiple UEs attempt to execute HO simultaneously.
This high correlation leads to many simultaneous requests from UEs (\textit{e.g.}, A3 events for multiple UEs occurring simultaneously), resulting in a high collision rate and prolonged access delays, negatively impacting network performance. To tackle this challenge, an efficient and effective HO protocol is needed to manage such a large volume of HO requests, minimize the collision rate and access delay, and ensure a high HO success rate.
To this end, we propose a novel HO protocol to address these issues, which will be discussed in the following.

%% file: Table/Table_NTNparameter_3GPP.tex
\begin{tabular}{l|r|r}
\toprule[1.0pt]
\textbf{Parameter} & \textbf{Value (Handheld)} & \textbf{Value (VSAT)} \\
\midrule[.5pt]
Carrier frequency, $f$ & 2\,[GHz] ($\mathrm{S}$-band) &  30\,[GHz] ($\mathrm{K_{a}}$-band) \\
Bandwidth, $B$ & 0.4\,[MHz] & 400\,[MHz]\\
TX transmit power & 200\,[mW] (23\,[dBm]) & 2\,[W] (33\,[dBm]) \\ 
TX antenna gain, $G_\mathrm{T}$ & 0\,[dBi] & 43.2\,[dBi] \\ 
\midrule[.5pt]
Atmospheric loss, $P_{\mathrm{A}}$ & 0.1\,[dB] & 0.5\,[dB]  \\
Shadowing margin, $P_{\mathrm{SM}}$ & 3\,[dB] & 0\,[dB]  \\
Scintillation loss, $P_{\mathrm{SL}}$ & 2.2\,[dB] & 0.3\,[dB]  \\
\midrule[.5pt]
$G/T$ & 1.1 [dB/K] & 13 [dB/K] \\ 
\midrule[.5pt]
Boltzmann constant, $k$ & \multicolumn{2}{r}{-228.6\,[dBW/K/Hz]} \\        
\bottomrule[1.0pt] 
\end{tabular}


%% file: Section/SEC4.tex
\begin{figure*}[!h]
\centering
\subfloat[Conventional HO protocol.]{
\includegraphics[width=0.8\columnwidth]{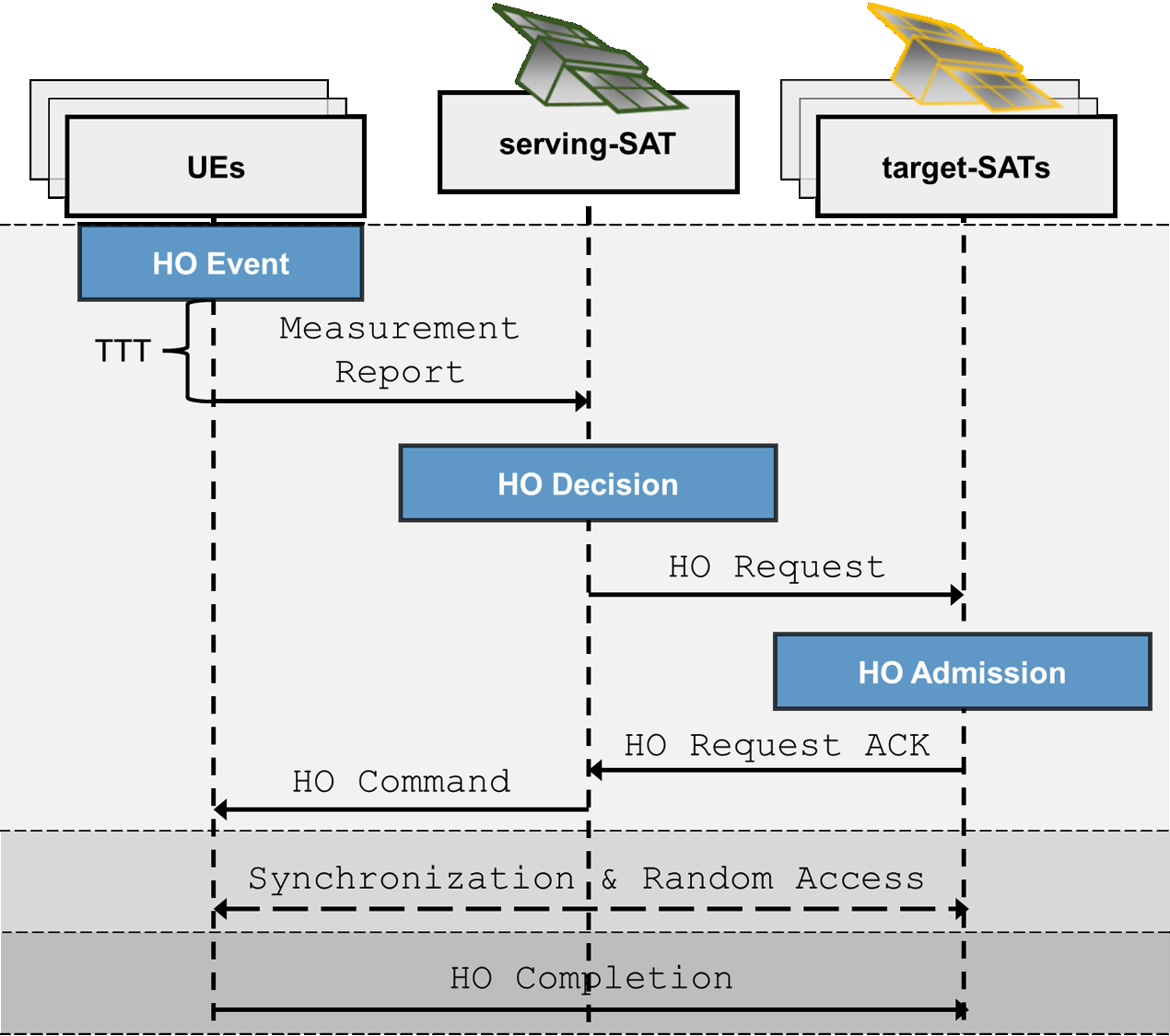}
\label{fig:existingHO}
}
\subfloat[Proposed DRL-based HO protocol (DHO).]{
\includegraphics[width=0.8\columnwidth]{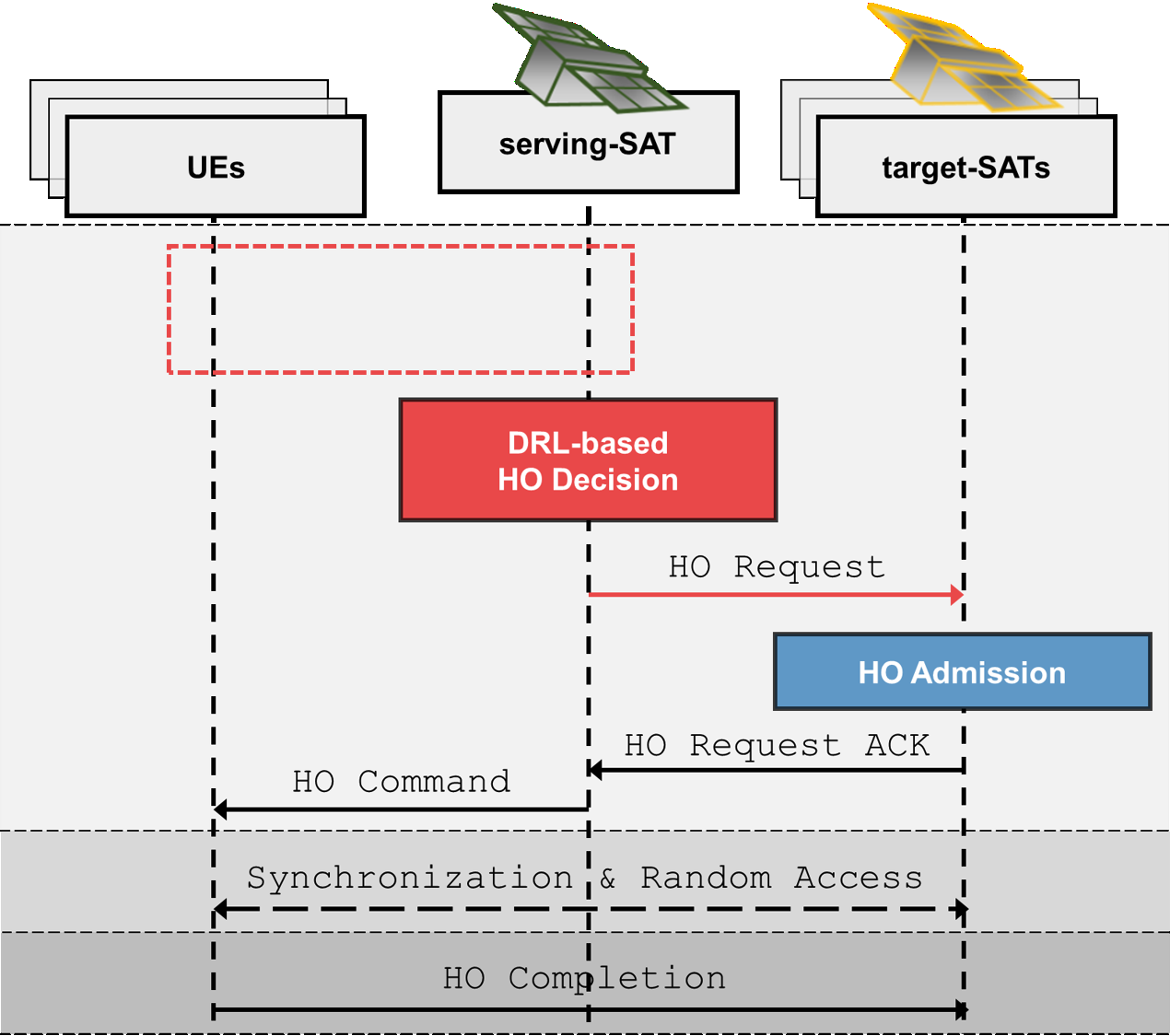}
\label{fig:proposedHO}
}
\\
\caption{The sequence diagrams of traditional and proposed HO protocols for LEO SAT networks.}
\label{fig:HOforLEOSAT}
\vspace{-1.em}
\end{figure*}

\tpurp{
Existing HO methods for LEO SAT networks are limited by long propagation delays and high power consumption for uplink signaling. To address these issues, we propose a DRL-based HO protocol in which the serving-gNB agent predicts the UE's network signal information and sends a \texttt{HO Request} to the target-gNB without the need for a \texttt{MR}. This simplifies the HO preparation phase and improves HO performance by overcoming long propagation delays while saving signaling power. Fig.~\ref{fig:HOforLEOSAT} illustrates the proposed protocol. In this section, we introduce the LEO SAT network scenario, evaluation metrics, and detailed protocol design.
}

\subsection{Network Scenario}

\BfPara{Configuration of LEO SATs and Ground UEs} 
Consider a set $\mathcal{K}$ of orbital planes around Earth. For each orbital plane $k \in \mathcal{K}$, there is a set $\mathcal{I}_k$ of LEO SATs orbiting on that plane. Additionally, there is a set $\mathcal{J}$ of UEs deployed on the ground\footnote{
Our primary focus is on fixed VAST-type UEs, but mobile handheld-type UEs can also be considered with minor adjustments.} inside an area $A$. The position of UE $j \in\mathcal{J}$ is expressed as a 3-dimensional real vector on Cartesian coordinates denoted by $\vb*{q}_{j} = (q^{x}_{j},q_{j}^{y},q_{j}^{z})\in \mathbb{R}^3$. Similarly, the position and velocity of SAT $i$ at time $t\geq 0$ are denoted by $\vb*{q}_{i}(t) = (q_{i}^{x}(t),q_{i}^{y}(t),q_{i}^{z}(t)) \in \mathbb{R}^{3}$ and $\vb*{v}_i(t) = (v_{i}^{x}(t),v_{i}^{y}(t),v_{i}^{z}(t)) \in \mathbb{R}^3$, respectively. The set of all SATs is denoted by $\bigcup_{k\in\mathcal{K}}\mathcal{I}_k$. We assume that the number of SATs on each orbital plane is equal to each other, given as $|\mathcal{I}_k| = I$ for all $k\in\mathcal{K}$, and all SATs are moving in a uniform circular motion with the same orbital period $T$ which is an acceptable approximation for the duration of visibility of the SAT to the UE. The arc length between any two neighboring SATs on the same orbital plane is also assumed to be equal to each other.

Consider that time is discretized in slots of length $\tau$. Let $\vb*{q}_{i}[0]$ be the initial position of the SAT $i \in \bigcup{k \in \mathcal{K}}\mathcal{I}_k$ at time $t = 0$. Then, by following the discrete-time state-space model \cite{UAV_SCA_YZ, P4C_JH}, the position of SAT $i$ at time $t = m\tau$ can be expressed as
\begin{align}
\vb*{q}_{i}[m] = \vb*{q}_{i}[0] + \tau \sum_{m' = 1}^{m}\vb*{v}_i[m'\tau]. \label{eq:C_LEO_q}
\end{align}

\BfPara{HO Procedure} 
Consider LEO SAT HO scenario, where UEs in the cell $\mathcal{O}$ connect to a serving-SAT, which is acting as a serving-gNB. There are other SATs in the field-of-view (FoV) that are candidates for the HO target, referred to as target-gNBs (or target-SATs), and the serving-SAT can initiate a HO request to the target-SATs if it estimates it necessary. Suppose that the source-SAT and target-SATs are the closest SATs on each orbital plane for the UEs in $\mathcal{A}$.
Suppose that there are $N$ HO opportunities during an interval $T$, after which a UE loses connectivity with the serving-SAT unless it successfully conducts a HO to a target-SAT. The time duration of each HO time slot is $\tau = \frac{T}{N}$, such that $\tau \in \mathbb{Z}^+$. For simplicity, we will focus only on the $N$ HO opportunities in the rest of this section, and suppose the first time slot starts at $t = 0$.
Here, the time duration of each $n$-th HO  is discretized with $\tau$, \textit{i.e.}, $t = n \tau, \forall n\in \{1,2,\dots,N\}$.

At each HO opportunity, each UE decides whether to send a \texttt{HO Request} and, if so, selects target-SATs to which it will send it.
Such a set of actions is represented by $\lbrace 0, 1, \ldots, K - 1 \rbrace$, where $K:=|\mathcal{K}|$ is the number of orbital planes and $k = 0$ corresponds to the serving-SAT.
The HO action of UE $j \in \mathcal{J}$ at the $n$-th HO opportunity is denoted by
\begin{equation}
a_{j}[n] \in \lbrace 0, 1, \ldots, K - 1 \rbrace. \label{action_HO}
\end{equation}
Note that $a_{j}[n] = 0$ means that the serving-SAT does not send a \texttt{HO Request} to any target-SAT at the $n$-th HO opportunity and waits for the next one.
Also, note that there can be no HO to another SAT on the same orbital plane.

\begin{figure}[!t]
\centering
\includegraphics[width=.9\linewidth]{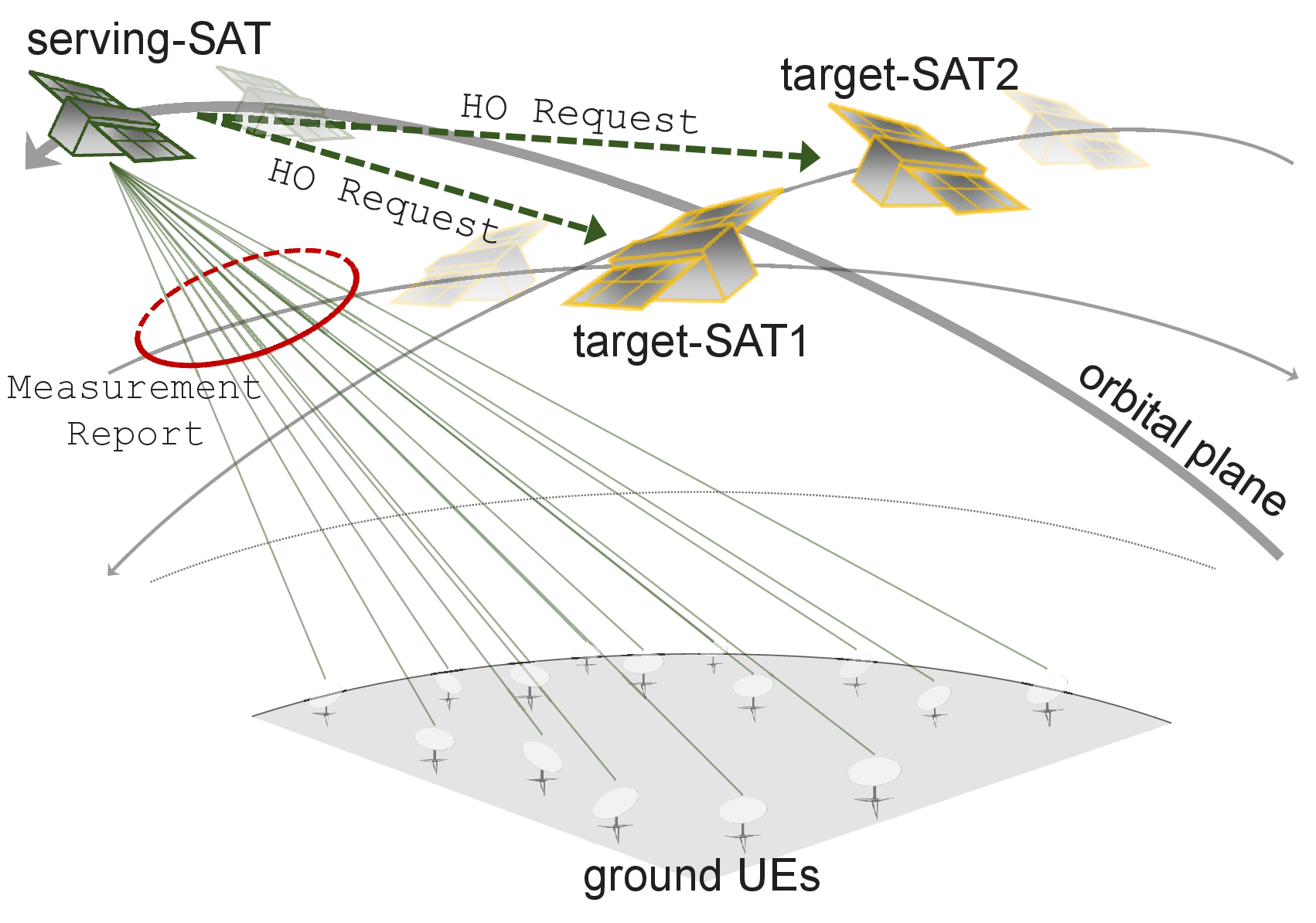}
\caption{Network scenario. UEs are required to perform HO, transitioning their connections from the serving-SAT to either target-SAT1 or target-SAT2.}
\label{fig:networkscenario}
\vspace{-1.em}
\end{figure}

\BfPara{Successful HO or Collision}
There can be two types of collisions in the LEO SAT HO scenario:
\begin{enumerate}
    \item \textit{negative acknowledgment (NACK) due to lack of resource blocks (RBs)}: If the number of received \texttt{HO Request} sent by serving-SAT is greater than the number of RBs in the target-SAT, collisions will occur.
    \item \textit{physical random access channel (PRACH) collision}: Collisions occur if multiple UEs that receive the \texttt{HO Command} attempt to access the same target-SAT with the same preamble signature.
\end{enumerate}
Note that once UEs receive the \texttt{HO Command}, the random access (RA) signaling is carried out in two steps, as specified in Release 16 of 5G-NR \cite{3GPP_NR_RACH_Rel16}. 
Specifically, UEs select a preamble from a set of available preambles for each target-SAT uniformly at random, represented by $p_{k}[n] \in \lbrace 1,2,\dots, P \rbrace, \ \forall k \in \lbrace 1, \ldots, K - 1 \rbrace$, where $P$ represents the number of resources that each SAT can grant during the data transmission duration.
In the first step of the RA signaling process, UEs that have determined to access send preambles to the corresponding SATs. 
In the second step, SATs send feedback to confirm whether there were collisions or not for each chosen preamble. UEs that have chosen colliding preambles fail to access, while those that have chosen preambles without collision succeed in completing the HO, \textit{i.e.}, \texttt{HO Completion}, which is denoted by a binary indicator $a^{\mathrm{HO}}_{j}[n]$. Specifically, $a^{\mathrm{HO}}_{j}[n]=1$ indicates the HO has been completed for UE $j$, while $a^{\mathrm{HO}}_{j}[n]=0$ indicates the HO has not been completed for UE $j$ at time slot $n$ and $a^{\mathrm{HO}}_{j}[0]=0, \ \forall j$.

\subsection{Evaluation}
The performance of HO in the LEO SAT networks can be evaluated mainly in terms of collision rate and access delay, as explained in the following.

\subsubsection{Collision Rate}
The collision rate is a measure of the proportion of unsuccessful HO attempts, which can occur due to two factors: \textit{NACK due to lack of RBs} and \textit{PRACH collision}.

\BfPara{Collision (\textit{NACK due to lack of RBs})} 
This type of collision occurs when there are not enough RBs available on the target-SAT to support the number of UEs attempting to access it.
We can represent each \texttt{HO Request} by defining the request indicator  $h^{\mathrm{R}}_{k, j}[n]$ for UE $j \in \mathcal{J}$ attempting to access target-SAT $k \in \mathcal{K}$ at  $n$:
\begin{align}
h^{\mathrm{R}}_{k, j}[n] = 
\begin{cases}
1 & a_{j}[n] > 0,~a^{\mathrm{HO}}_{j}[n] = 0, \\
0 & \text{otherwise},
\end{cases} \ \forall j \in \mathcal{J}, k \in \mathcal{K}.
\end{align}

We can then calculate the collision rate $C^{\mathrm{R}}_{k}[n]$ for this type of collision as the ensemble proportion of UEs that attempted to access a target-SAT $k \in \mathcal{K}\backslash \{k\!=\!0\}$  with insufficient RBs:
\begin{align}
C^{\mathrm{R}}_{k}[n] =
\begin{cases}
0 \!&\! \mathrm{R}_{k}[n] - \sum^{J}_{j=1}{h^{\mathrm{R}}_{k, j}[n]} > 0, \\
\frac{\sum^{J}_{j=1}{h^{\mathrm{R}}_{k, j}[n]} - \mathrm{R}_{k}[n]}{J} \!&\! \text{otherwise},
\end{cases}
\end{align}
where $\mathrm{R}_{k}[n]$ represents the available RBs in target-SAT $k$ at time slot $n$, and $c^{\mathrm{R}}_{j}[n]$ indicates the collision due to insufficient RBs for UE $j\in\mathcal{J}$ at time slot $n$. 
In the event of a collision due to insufficient RBs in the target-SAT, the UE that will receive the HO Request acknowledgment (ACK) is selected randomly based on the availability of the RBs.
To identify the UE that receives the \texttt{HO Command}, we define the command indicator $h^{\mathrm{C}}_{j}[n] \in \lbrace 0, 1, \ldots, K - 1 \rbrace$. $h^{\mathrm{C}}_{j}[n]=0$ indicates that the UE has not received any \texttt{HO Command}, while $h^{\mathrm{C}}_{j}[n]=k$ indicates that the UE has received the command for target-SAT $k$.

\BfPara{Collision (\textit{PRACH collision})}
This is another type of collision that occurs when multiple UEs attempt to access the same SAT with the same preamble signature.

Denote the collision indicator of UE $j\in\mathcal{J}$'s RA at time slot $n \in \{1,\dots,N\}$ by $c^{\mathrm{P}}_{j}[n]$, and define it as
\begin{align}
c^{\mathrm{P}}_{j}[n] \!=\!
\begin{cases}
    1 & (h^{\mathrm{C}}_{j}[n],p_{j}[n]) = (h^{\mathrm{C}}_{j^{'}}[n],p_{j'}[n]),~a^{\mathrm{HO}}_{j}[n] = 0, \\
    0 & \text{otherwise},
\end{cases}
\end{align}
if $a_j[n] \neq 0$, and otherwise we have $c^{\mathrm{P}}_j[n] = 0$.

The collision rate $C^{\mathrm{P}}[n]$ is defined as the ensemble proportion of UEs that experienced a PRACH collision and is calculated by averaging $c^{\mathrm{P}}_{j}[n]$ over all UEs in $\mathcal{J}$, which is defined as 
\begin{align}
C^{\mathrm{P}}[n] = \frac{1}{|\mathcal{J}|} \sum_{j \in \mathcal{J}}{c^{\mathrm{P}}_{j}[n]}.
\end{align}

Finally, the \emph{average collision rate} $C[n]$ is calculated by summing the collision rate for the lack of RBs $C_{k}^{\mathrm{R}}[n]$ over all target-SATs $k \in \mathcal{K}\backslash \{k\!=\!0\}$ and adding the PRACH collision rate $C^{\mathrm{P}}[n]$:
\begin{align}
C[n] = \sum^{K-1}_{k=1} C_{k}^{\mathrm{R}}[n] + C^{\mathrm{P}}[n]. \label{collision}
\end{align}
Note that a UE cannot experience both types of collisions (NACK due to lack of RBs and PRACH collisions), simultaneously. This is due to the fact that the UE initiates the RA process only after receiving the HO Command, which is sent by the serving-SAT upon receiving the HO Request ACK from the target-SAT. Thus, the RA process can only be initialized when there is no collision in terms of NACK due to lack of RBs.

\subsubsection{Access Delay}
Access delay is another metric in HO scenarios that measures the time it takes for a UE to successfully HO to a target-SAT. The access delay is calculated as the average time it takes for a UE to successfully access a target-SAT, taking into account the number of failed attempts and the time duration between each time slot.


The \emph{average access delay} is then calculated as:
\begin{align}
D[n] = \frac{1}{|\mathcal{J}|}\sum_{j\in\mathcal{J}} (1 - a^{\mathrm{HO}}_{j}[n]), \label{AccessDelay}
\end{align}
This calculation takes into account the total number of failed HO and the time duration between each time slot to provide an accurate measure of the average access delay.

\subsubsection{Successful HO (Access) Rate}
The successful HO rate is a metric that measures the proportion of UEs that successfully hand over to a target-SAT among all UEs that attempted to do so. It is calculated as the average of the access indicator of all UEs:
\begin{align}
H = \frac{1}{|\mathcal{J}|} \sum_{j\in\mathcal{J}} a^{\mathrm{HO}}_{j}[N].
\end{align}

\subsection{Protocol Design}
In this section, we propose a novel HO protocol for LEO SAT networks, called DHO. As the DHO utilizes DRL techniques for training, we first formulate the problem and cast the problem into a Markov decision process (MDP) model.

\subsubsection{Sequence of DHO}
The proposed DHO protocol for LEO SAT networks includes the following sequence of steps:
\begin{enumerate}
  \item \texttt{HO Decision}: serving-SAT makes a decision for \texttt{HO Request}, including target-SAT selection and backoff for each UE.
  \item \texttt{HO Admission}: target-SAT sends \texttt{HO Request ACK} if it has available RBs to support the HO request. The serving-SAT then sends \texttt{HO Command} to the UEs that are permitted to hand over to the target-SAT.
  \item \texttt{Random Access}: UEs that receive \texttt{HO Command} attempt to access the target-SAT designated by the serving-SAT, by randomly choosing preambles for RACH access.
  \item \texttt{HO Completion}: Completing HO if the UE successfully transfers to the target-SAT specified by the serving-SAT.
\end{enumerate}
Notably, the proposed DHO protocol simplifies the process by eliminating the need for the \texttt{MR}.

To achieve our objective, we employ the DRL approach, driven by two crucial factors:
Firstly, the DRL algorithm has the ability to handle complex, high-dimensional decision-making problems, making it a suitable choice for optimizing HO procedure in LEO SAT networks. Secondly, the environment in LEO SAT networks is dynamic and stochastic, characterized by rapidly changing network conditions. The DRL algorithm is capable of adapting to these fluctuations and making decisions in real-time, whereas other optimization techniques, such as convex optimization or dynamic programming, may not be equipped to handle such a complex and unpredictable environment.

\subsubsection{MDP Modeling}
The optimization problem mathematically formulated in DHO is to minimize both access delay and collision rate while taking into account the constraints of LEO SAT networks. The problem can be represented as:
\begin{align}
\min_{ \scriptsize \begin{array}{c} \scriptsize a_{j}[n] \end{array} } \sum_{n=1}^{N} D[n] + \nu C[n], \ \textrm{s.t.} \ \eqref{eq:C_LEO_q}. \label{eq:OptObj}
\end{align}
where $\nu$ is a normalization coefficient that balances the trade-off between access delay and collision rate, and $N$ is the period of time during which HO may occur. 

To recast the problem in \eqref{eq:OptObj} as a DRL problem, in the following, we first model the network scenario of LEO SAT-based NTN as an MDP.

\BfPara{Environment}
\emph{Environment} consists of ground UEs, serving-SAT, and target-SATs interacting with each other, which follows an MDP model. At each time step $n$, the serving-SAT is an agent that observes a state $\mathbf{s}[n]\in\mathcal{S}$, and takes action $\mathbf{a}[n]\in\mathcal{A}$ based on a state-action policy $\pi$. 
Given this action, the state of the agent transitions to $\mathbf{s}[n+1]$, and in return, the agent receives a reward $r[n]$ that reinforces following an optimal policy $\pi^*$.

\BfPara{State}
In the MDP model, the state at time index $n$ is defined as:
\begin{align}
\mathbf{s}[n] &= \lbrace n, \vb*a^{\mathrm{HO}}[n], \mathbf{a}[n-1] \rbrace, \label{eq:state}
\end{align}
where $n$ represents the time index, $\vb*a^{\mathrm{HO}}[n] = \{ a_{1}^{\mathrm{HO}}[n], a_{2}^{\mathrm{HO}}[n],\cdots, a_{J}^{\mathrm{HO}}[n] \}$, and $\mathbf{a}[n-1]$ is the previous action taken by the UEs. 
The inclusion of the time index and previous action serves as a \emph{fingerprint} for stabilizing experience replay in the DRL process. Besides, the binary indicator of accessed UEs provides information on the current state of the network.
Note that $\mathbf{s}[0]$ represents the state for $n=0$, which is initialized randomly.

The state information is designed to be minimal while still providing sufficient information for decision-making. Each piece of information in the state is carefully selected through extensive tests, as shown in our ablation study in Appendix \ref{sec:appendix}. It is worth noting that our proposed DHO utilizes the locally-observable information and does not require additional information, such as the position of the SATs or others, as the accessed UEs provide sufficient information for DHO training. This minimal state design not only reduces data collection overhead but also enhances the convergence of DRL training.

\BfPara{Action}
The action space $\mathcal{A}$ in our environment pertains to HO decisions.
In order to determine the \texttt{HO Request} for each UE at each HO opportunity, our agent, representing the serving-SAT, employs the HO action $a_{j}[n]$ as defined in \eqref{action_HO}.

To define the set of HO actions, $\mathcal{A}$, we utilize a one-hot encoded representation. Specifically, we define the set of HO actions for UE $j$ at time index $n$ as:
\begin{equation}
\mathbf{a}_j[n]=\{a_0,a_1,a_2,\cdots,a_{K-1}\}, \ \mathrm{s.t.} \sum_{k=0}^{K-1} a_k=1,    
\end{equation} 
where $a_k$ for $k\neq 0$ denotes the association with the $k$-th orbital plane, and $a_0$ implies that the serving-SAT does not send a \texttt{HO Request} to any target-SAT at the $n$-th HO opportunity and instead waits for the next one.
This one-hot encoded representation ensures that exactly one action is selected at each time step, and allows for a clear and efficient representation of the HO decisions made by the agent.
Thus, the complete set of action is given as:
\begin{align}
\mathbf{a}[n] &= \begin{bmatrix}
       \mathbf{a}_1[n] \\
       \mathbf{a}_2[n] \\
       \vdots \\
       \mathbf{a}_J[n]
     \end{bmatrix}.
\end{align}

\BfPara{Reward}
The reward function in our proposed DHO is designed to reinforce the serving-SAT to make optimal access decisions by penalizing access delay and collision rate.
The reward function, $r[n]$, is defined as the negative of the objective function in \eqref{eq:OptObj}, which captures this goal:
\begin{align}
r[n] = - D[n] - \nu C[n]. \label{eq:reward}
\end{align} 
Here, the normalization coefficient $\nu$ balances the trade-off between access delay and collision rate, which will be discussed in Sec. \ref{sec:tradeoff}, specifically in Table \ref{table:coeffRatio_Reward}.

It is worth noting that the state and reward information used in our MDP model is locally observable, making it possible to train and execute the DHO in a fully distributed manner. This means that the regenerative type SATs can easily implement our proposed DHO.

\subsection{Algorithm Details} \label{sec:algorithmDetail}
\tpurp{
The proposed DHO employs IMPALA~\cite{espeholt2018impala}, which is a DRL algorithm that provides several advantages over other DRL algorithms, such as DQN \cite{DRL_DQN}, A3C \cite{DRL_A3C}, and PPO \cite{DRL_PPO}\footnote{
DHOs can also be trained with other DRL algorithms that support multi-discrete actions.
}. IMPALA is an off-policy reinforcement learning algorithm that utilizes parallel actor-learners and importance sampling to improve sampling efficiency and scalability. 

In off-policy training, the variance is largely due to the difference between the policy that is used to generate the behavior (behavior policy) and the policy that is being improved (target policy). 
IMPALA addresses the policy mismatch issue in off-policy learning by using \emph{V-trace} targets, which are estimated using truncated importance sampling. V-trace targets correct the value estimate by incorporating the difference between the behavior policy and the target policy, which can significantly improve learning performance.
IMPALA provides scalability and improved sample efficiency by employing asynchronous updates and importance sampling. It allows for efficient parallelization and effective utilization of collected data, striking a balance between exploration and exploitation. These advantageous features make IMPALA particularly suitable for large-scale environments and complex tasks, such as the HO process in LEO SAT networks.
}

The process of how the learner updates its policy in IMPALA is as follows:
\begin{enumerate}
\item Initialize and update the actor's policy based on the current learner's policy.
\item The actor collects experience in environments.
\item The actor delivers the collected experience and its policy to the learner.
\item The learner trains its policy based on the experience and the actor's policy it receives.
\end{enumerate}

In IMPALA, which is an asynchronous distributed training model, even if different actors update their policies based on the same learner, they may diverge due to the inherent randomness and asynchrony of the training process.

\subsubsection{V-trace}
In the context of discounted infinite-horizon RL in MDP, we aim to find a policy $\pi$ that maximizes the expected sum of future discounted rewards. The value function $V^{\pi}(\mathbf{s})$ represents the expected cumulative rewards starting from state $\mathbf{s}$, and it is defined as $ E_{\pi} \left[ \sum_{n=1}^{N}{\gamma^t r[n]} \right]$, where $\gamma$ is the discount factor in the range $\left[ 0,1 \right)$.

Consider a trajectory $\left(\mathbf{s}[n], \mathbf{a}[n], r[n]]\right)_{n=s}^{n=s+k}$ generated by the actor following a policy $\mu$. The $k$-steps \emph{V-trace target} $v_n$ for approximating the value at state $\mathbf{s}[n]$, is defined as:
\begin{equation}
v[n] = \textstyle{V(\mathbf{s}[n]) + \sum_{n=s}^{s+k-1} \gamma[n-s] \left( \prod_{i=s}^{n-1} c[i] \right) \delta_n V}, \label{eq:target.off}
\end{equation}
where $\delta_n V = \rho[n] \left(r[n] + \gamma V(\mathbf{s}[n+1])-V(\mathbf{s}[n])\right)$ is the temporal difference for $V$. The terms $\rho[n]$ and $c[n]$ are truncated importance sampling (IS) weights, with $\rho[n]=\min\left(\bar\rho, \frac{\pi(\mathbf{a}[n]|\mathbf{s}[n])}{\mu(\mathbf{a}[n]|\mathbf{s}[n])}\right)$ and $c[n]=\min\left(\bar c, \frac{\pi(\mathbf{a}[n]|\mathbf{s}[n])}{\mu(\mathbf{a}[n]|\mathbf{s}[n])}\right)$. It is assumed that the truncation levels satisfy $\bar \rho\geq \bar c$.
The truncation levels $\bar c$ and $\bar \rho$ play different roles in the algorithm: $\bar \rho$ impacts the nature of the value function we converge to, whereas $\bar c$ impacts the speed at which we converge to this function.


Off-policy learning is important in the decoupled distributed actor-learner architecture because of the lag between when actions are generated by the actors and when the learner estimates the gradient. By incorporating the V-Trace target, IMPALA addresses the challenge of policy mismatch (policy-lag) in distributed environments, allowing for a more accurate estimation of the advantage function and improved learning performance.

\subsubsection{V-Trace Actor-Critic algorithm}
Under off-policy settings, we can use the IS weight between the policy being evaluated $\pi_{\bar\rho}$ and the behavior policy $\mu$ to update our policy parameter in the direction of $$\mathbb{E}_{\mathbf{a}[n]\sim\mu(\cdot|\mathbf{s}[n])}\bigg[
\frac{\pi_{\bar{\rho}}(\mathbf{a}[n]|\mathbf{s}[n])}{\mu(\mathbf{a}[n]|\mathbf{s}[n])}
\cdot\nabla\log\pi_{\bar{\rho}}(\mathbf{a}[n]|\mathbf{s}[n]) q[n]|\mathbf{s}[n]\bigg].$$
Here, $q[n]:=r[n]+\gamma v[n+1]$ is an approximation of $Q^{\pi_{\boldsymbol{\bar{\rho}}}}(\mathbf{s}[n],\mathbf{a}[n])$, built from the V-trace estimate $v_{s+1}$ at the next state $x_{s+1}$.

Consider a parametric representation of the value function $V_{\phi}$ and the current policy $\pi_{\theta}$. Trajectories are generated by actors following a behavior policy $\mu$. The V-trace targets $v[n]$ are defined by the equation \eqref{eq:target.off}.
During training, the value parameters $\phi$ are updated using gradient descent on the $l2$ loss towards the target $v[n]$. This update is performed in the direction of 
$$\big( v[n] - V_\phi(\mathbf{s}[n])\big) \nabla_\phi V_\phi(\mathbf{s}[n]).$$ 
The policy parameters $\theta$ are updated in the direction of the policy gradient, which is given by $$\rho[n] \nabla_\theta\log\pi_\theta(\mathbf{a}[n]|\mathbf{s}[n]) \big( r[n]+\gamma v[n+1] - V_\phi(\mathbf{s}[n])\big).$$ 
To prevent premature convergence, an entropy bonus, similar to A3C, can be added in the direction $$-\nabla_\theta \sum_{\mathbf{a}} \pi_\theta(\mathbf{a}|\mathbf{s}[n]) \log\pi_\theta(\mathbf{a}|\mathbf{s}[n]).$$


The detailed process of the training is described in Algorithm~\ref{alg:IMPALA}.

\begin{algorithm}[!t]
\small
    Initialize the \textit{learner} network's weights $\boldsymbol{\phi}$ $\rightarrow$ $\boldsymbol{\phi}_0$\,;\\
    \For{Epoch = 1, MaxEpoch}{
        \For{each \textit{actor} $i$, $\forall i \in [1,I]$}{
        $\triangleright$ \textbf{Initialize LEO SAT Environments}:\,\,\,\,\, \ \
        Set replay buffer $\mathcal{D}_i=\{\}$ and $\mathbf{s}_i[n]$ $\rightarrow$ $\mathbf{s}_i[0]\,;$ \\

        $\triangleright$ Update \textit{actor} weight $\boldsymbol{\theta}_i$ from the \textit{learner}: Download weights $\boldsymbol{\theta}_i$ = $\boldsymbol{\phi}\,;$ \\
        
        \For{$\text{Step} = 1, \text{MaxStep}$}{    
            $\triangleright$ Select the action $\mathbf{a}_i[n]$ based on its policy $\mu_{\boldsymbol{\theta}_i}(\mathbf{a}_i[n]\,|\,\mathbf{s}_i[n])$ at every step$\,;$ \\
            $\triangleright$ Get reward $r_i[n]$ and move $\mathbf{s}_i[n] \rightarrow \mathbf{s}_i[n+1]\,;$ \\
            $\triangleright$ Set experience $\xi_i[n]= \{\mathbf{s}_i[n], \mathbf{a}_i[n], r_i[n], \mathbf{s}_i[n+1]\}\,;$ \\
            $\triangleright$ Update replay buffer: $\mathcal{D}_i=\mathcal{D}_i\cup\xi_i[n]\,;$ \\
            }
        $\triangleright$ Upload $\mu_{\boldsymbol{\theta}_i}$ and $\mathcal{D}_i$ to the \textit{learner}$\,;$ \\
        $\triangleright$ Update weights $\boldsymbol{\phi}$ of the \textit{learner} network $\pi_{\boldsymbol{\phi}}$ on the basis of $\mu_{\boldsymbol{\theta}_i}$ and $\mathcal{D}_i\,;$
        }
    }
\caption{Training process for the proposed DHO}
\label{alg:IMPALA}
\end{algorithm}

\subsection{Complexity Analysis} \label{sec:complexity}
In the complexity analysis, we first discuss the training complexity of the DHO, which utilizes the IMPALA algorithm for training. We then introduce the execution complexity of our proposed DHO compared to the conventional HO protocol.

\BfPara{Training complexity}
Policy-based DRL, utilizing policy gradient, has been demonstrated to possess superior reward convergence compared to value-based DRL~\cite{sutton1999policy}. One example of policy-based DRL is the actor-critic algorithm, such as A2C and A3C. However, as the actor-critic network is an \emph{on-policy} method, it lacks the ability to utilize an experience replay buffer, as seen in off-policy training methods like DQN~\cite{mnih2013playing}. This lack of experience replay buffer can lead to a higher susceptibility to converge to a local minimum or experience oscillations due to correlations between training samples and rapidly changing data distributions.

\tpurp{
IMPALA, on the other hand, leverages \emph{off-policy} multiple actors-learner distributed learning with V-Trace to provide a faster learning rate and improved performance. The use of multiple actors and a distributed learning approach also allows for the efficient utilization of computing resources, making faster convergence possible. 

After training, our DRL-based HO agent, DHO, can output an action (\textit{i.e.}, \texttt{HO Request}) in less than a few milliseconds on NVIDIA GeForce RTX 3080 Ti GPU. Further details of the comparison and convergence study of the IMPALA DRL algorithm can be found in Appendix \ref{sec:appendixB}.
}

\BfPara{Execution complexity}
One of the key advantages of the proposed DHO is its simplification of the HO procedure by skipping the \texttt{MR} step. This simplification reduces the overall HO processing time and conserves UL signaling power that would otherwise be used for the \texttt{MR} in conventional HO protocols. In the DHO, the serving-SAT agent is capable of directly sending a \texttt{HO Request} to the target-SATs without relying on the \texttt{MR}. This is achieved by leveraging the serving-SAT agent's learned understanding and prediction of the UE's channel condition. The elimination of the \texttt{MR} step significantly reduces the execution complexity, enhancing the efficiency and effectiveness of the proposed DHO compared to the conventional HO protocol.

%% file: Section/SEC5.tex
In this section, we demonstrate the effectiveness of the proposed DHO for LEO SAT networks by evaluating the performance metrics of access delay and collision rate.

\begin{table}[h!]   
\centering
\resizebox{1\linewidth}{!}
{\begin{minipage}[t]{0.5\textwidth}
\caption{Simulation parameters.}
\centering
\scriptsize
\input{Table_Result/Table_Parameter.tex}

\label{table:Paramter}
\end{minipage}}
\end{table} 

\subsection{Simulation Setup}

\begin{figure*}[h!]
\centering
\includegraphics[width=0.45\linewidth]{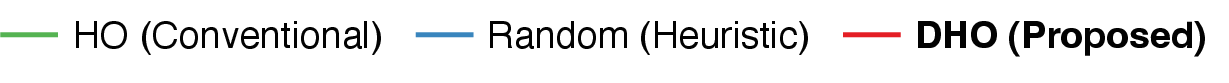} \vspace{-1.5em}\\
\subfloat[Average access delay \label{fig:comparion_RB_AD}]{\includegraphics[width=0.33\linewidth]{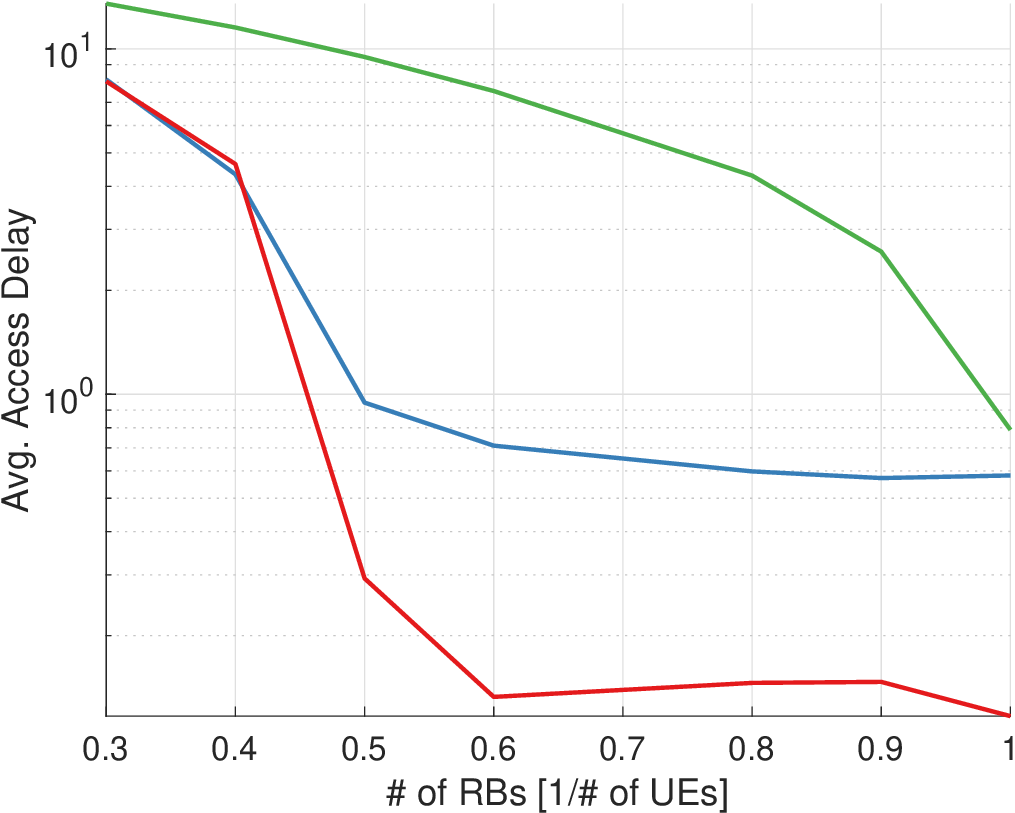}}
\subfloat[Average collision rate (NACK due to lack of RBs) \label{fig:comparion_RB_CR}]{\includegraphics[width=0.33\linewidth]{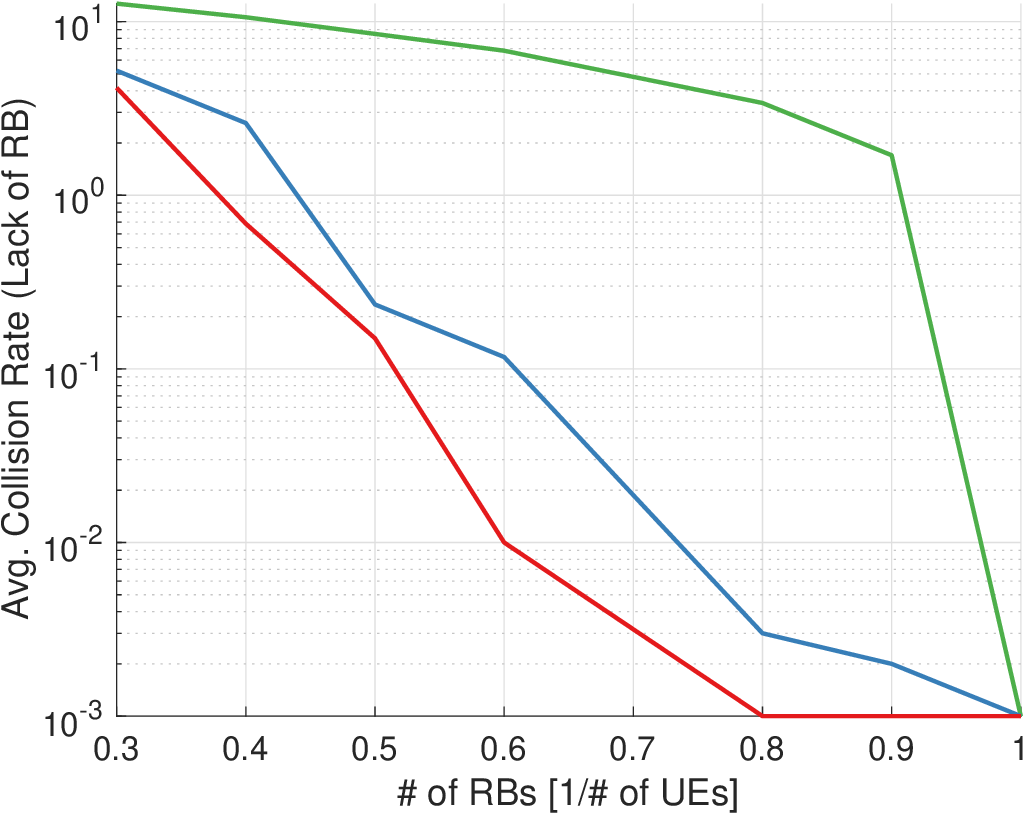}}
\subfloat[Average collision rate \label{fig:comparion_CR}]{\includegraphics[width=0.33\linewidth]{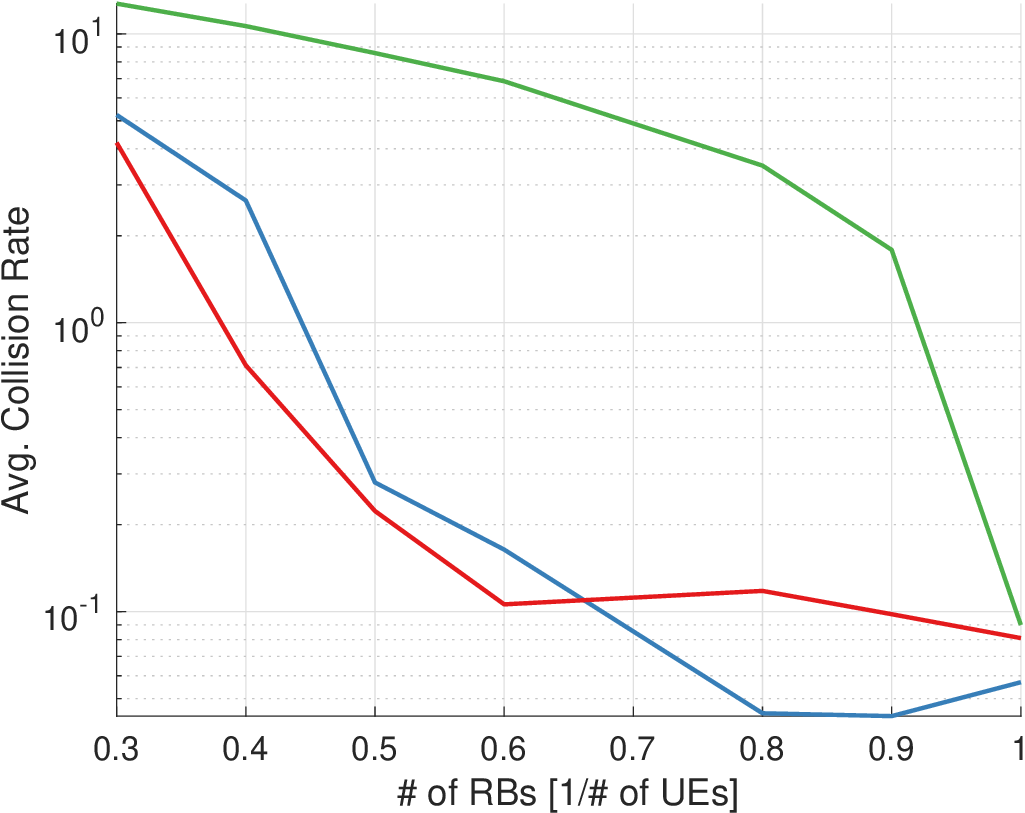}}
\vspace{-1.em}
\caption{Impact of the number of RBs. $x$-axis represents the number of RBs per UE  on each target-SAT, \textit{i.e.}, $R_k/J$  ($J=10$ and $P=5 J$).}
\label{fig:comparion_RB}
\vspace{-.5em}
\end{figure*}

\begin{table*}[h!]
\centering          
\resizebox{1.25\columnwidth}{!}{\begin{minipage}[h]{1.35\columnwidth}
\centering
\input{Table_Result/Table_RB.tex}
\end{minipage}}
\caption{Comparison of access delay and collision rate of DHO with benchmark methods (HO, Random) for the number of RBs in each target-SAT ($J=10$ and $P=5 J$).}
\label{table:comparison_RB}
\vspace{-1.5em}
\end{table*}

\subsubsection{Environment} 
\tpurp{
In this study, unless otherwise specified, the ground UEs are uniformly distributed within a $1000 \times 1000$ [m$^2$] area. The scenario includes one serving-SAT ($k=0$) and two target-SATs ($k=1~\mathrm{or}~2$), orbiting at an altitude of $550$ [km] in three different orbital planes.
Given that LEO forms a spherical shape with its center corresponding with that of the Earth, the orbital speed of LEO SATs is determined based on their altitude using Kepler's law. 
The FoV of the UEs in a specific area of interest ($\mathcal{A}$) is assumed to cover one SAT for each orbital lane.
}

The parameters for conventional HO decision-making are as follows: 
The filtered HO measurement, $M_{\mathrm{L_1}}[n]$ (\textit{i.e.}, received signal strength indicator (RSSI)), is updated every HO measurement period ($T_{\mathrm{M}}$) at the UE. The measurement $M_{\mathrm{L_3}}[n]$ is evaluated by L3 infinite impulse response (IIR) filtering of L1 received power of downlink reference signals (RSRP) measurements for each HO decision update period ($T_{\mathrm{U}} = T_{\mathrm{M}} / \beta_{\mathrm{L_3}}$) with a forgetting factor $\beta_{\mathrm{L_3}} = 1/2^{(k_{\mathrm{IIR}} / 4)}$, where $k_{\mathrm{IIR}}$ is the IIR filter order. The L3 filtered measurement is given as \cite{Handover_3GPP_LTE_Parameter_2, Handover_3GPP_LTE_Parameter}: 
\begin{align}
M_{\mathrm{L_3}}[n] = \beta_{\mathrm{L_3}} M_{\mathrm{L_1}}[n] + (1 - \beta_{\mathrm{L_3}}) M_{\mathrm{L_3}}[n-1].
\end{align}
Note that in scenarios with high doppler shift and high speed, such as in the LEO SAT networks, the log-normal shadowing samples may not be highly correlated. In such cases, a shorter filtering period may result in more accurate HO decision-making.
When the event A3 criterion is met, \textit{i.e.}, the L3 filtered RSRP of the target cell exceeds that of the serving cell by a predetermined hysteresis margin (referred to as the event A3 offset), the UE sends a notification to the serving cell and relays this event A3 condition through a \texttt{MR}, thereby initiating the HO preparation process.

Table \ref{table:Paramter} presents a summary of the main parameters for the environment, the network scenario, specifically those related to HO, and the DRL training.

\begin{figure*}[h!]
\centering
\includegraphics[width=0.45\linewidth]{Figure/description.eps} \vspace{-1.5em}\\
\subfloat[Average access delay. \label{fig:comparion_PRACH_AD}]{\includegraphics[width=0.33\linewidth]{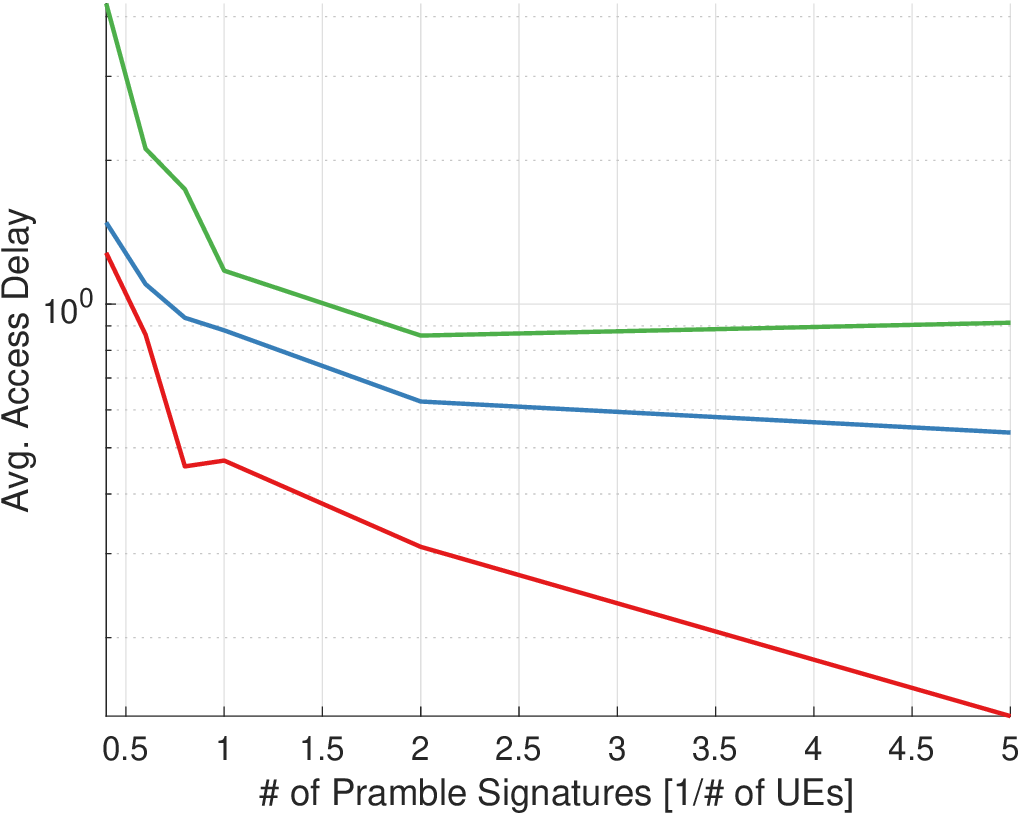}}
\subfloat[Average collision rate (PRACH collision) \label{fig:comparion_PRACH_CR}]{\includegraphics[width=0.33\linewidth]{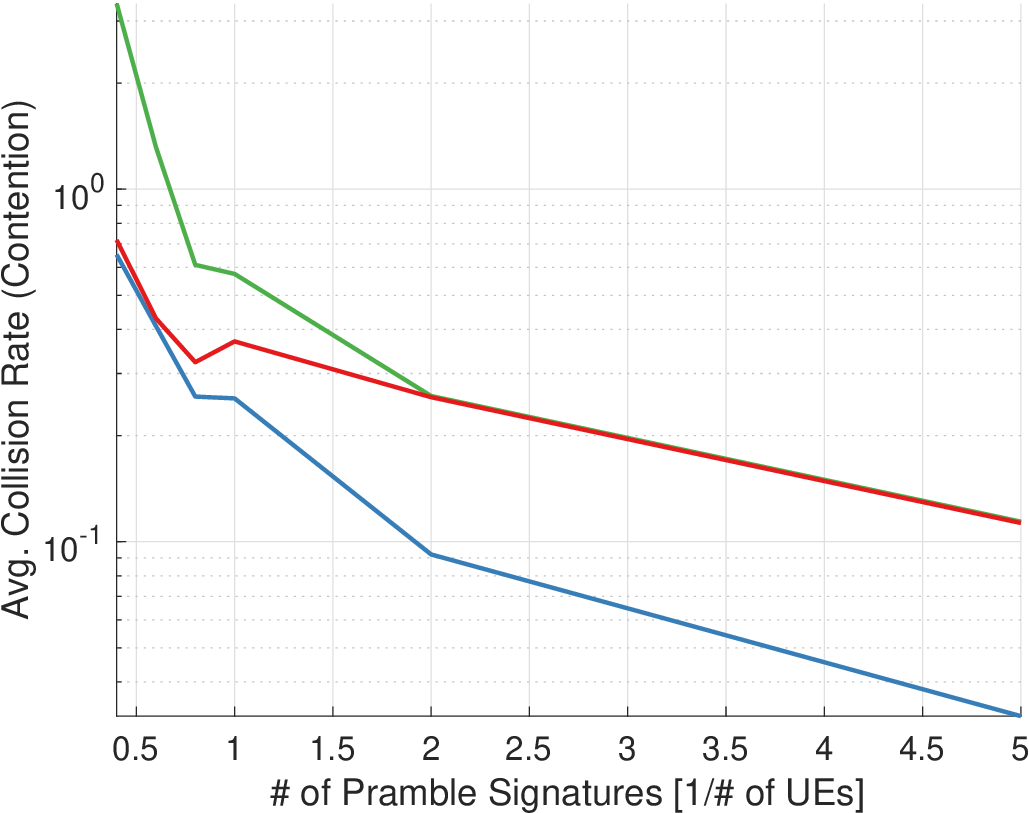}}
\caption{Impact of the number of preamble signatures. $x$-axis represents the number of preambles per UE  on each target-SAT, \textit{i.e.}, $P/J$ ($J=10$ and $R_k = J, \ \forall k$).}
\label{fig:comparion_PRACH}
\end{figure*}

\begin{table*}[!ht]
\centering          
\resizebox{1.0\columnwidth}{!}{\begin{minipage}[h]{1.08\columnwidth}
\centering
\input{Table_Result/Table_PRACH.tex}
\end{minipage}}
\caption{Comparison of access delay and collision rate of DHO with benchmark methods (HO, Random) for the number of preamble signatures for each target-SAT ($J=10$ and $R_k = J, \ \forall k$).}
\label{table:comparison_PRACH}
\vspace{-1.5em}
\end{table*}

\subsubsection{Benchmark}

Throughout this section, we evaluate the performance of two benchmark HO methods and our proposed LEO SAT networks-oriented HO, referred to as DHO, as outlined below:
\begin{enumerate}

\item \tpurp{\textbf{HO (Conventional)} protocol is a traditional method employed in 4G-LTE and 5G-NR networks. This method utilizes the A3 event to trigger HO\footnote{\tpurp{Here, the A3 event is used as it is widely used and has a higher occurrence probability in practical deployments. Other entering conditions, \textit{\textit{e.g.}}, A2, A4, or A5, can also be used for the conventional HO  (see Table \ref{table:A1_5}).}}, and the ground UE performs HO measurements using the estimated RSRP, while also applying filters to mitigate the impact of fading and measurement inaccuracies. At the start of the HO opportunity, the serving-SAT selects a target-SAT based on favorable RSRP. Additional details on this process can be found in Sec. \ref{sec:HO_overview} and in Table \ref{table:Paramter}, which lists the relevant parameters.
}

\item \textbf{Random (Heuristic)} protocol selects actions randomly for each HO opportunity in the environment. As a result, each UE receives a \texttt{HO Command} for a specific target-SAT or none at all. However, collisions may occur if multiple UEs try to access the same target-SAT with the same preamble signature (PRACH) when there is an insufficient amount of RB available on the target SAT. Here, the two-step RACH process is assumed for RA signaling, following the guidelines of Release 16 of 5G-NR \cite{3GPP_NR_RACH_Rel16}.



\item \textbf{DHO (Proposed)} is our proposed HO method trained using the IMPALA framework, as presented in Algorithm~\ref{alg:IMPALA}. This approach enables each serving-SAT to select the optimal HO action (\textit{e.g.}, \texttt{HO Request}) for all ground UEs to be sent to target-SATs. 
This method operates under partial state observability, whereby each serving-SAT agent interacts only with ground UEs and utilizes observable state information without communicating with other entities. 
The learning process can be seen as a partially observable MDP (POMDP).
\end{enumerate}

As discussed in Sec. \ref{sec:background}, the challenge of massive HO arises in LEO SAT networks due to the high density of ground UEs that need to HO simultaneously; besides, the fast orbital movement of LEO SATs leads to more frequent HO. Therefore, to evaluate the performance of LEO SAT networks under these conditions, we examine our proposed techniques based on two critical factors for a massive HO scenario: the number of available RBs and the number of preamble signatures.

\begin{table*}[!h]
    \centering          
    \caption{DHO agent behavior for various scenarios ($J=10$).}
    \resizebox{1.6\columnwidth}{!}{\begin{minipage}[h]{1.67\columnwidth}
    \centering
    \label{table:agentPolicy}
    \input{Table_Result/Table_AgentPolicy.tex}
    \end{minipage}}
\end{table*}

\begin{table*}[!h]
    \centering          
\caption{Trade-off between access delay and collision rate ($J=10$, $R_k = J, \ \forall k$, and $P=5 J$).}
    \resizebox{1.25\columnwidth}{!}{\begin{minipage}[]{1.26\columnwidth}
    \centering
    \label{table:coeffRatio_Reward}
    \input{Table_Result/Table_Objective.tex}
    \end{minipage}}
    \vspace{-1.em}
\end{table*}

\subsection{HO Performance Analysis}

\subsubsection{Impact of Number of RBs}
The impact of the number of available RBs in each target-SAT on access delay and collision rate is shown in Fig. \ref{fig:comparion_RB}.
The $x$-axis represents the ratio of the number of RBs in a target-SAT to the number of ground UEs, indicating the proportion of available RBs that each target-SAT can be allocateed to ground UEs.
It is important to note that in order to express the massive HO situation more efficiently, we consider the ratio of the number of resources per UE rather than directly considering the number of UEs and resources.

Specifically, Fig. \ref{fig:comparion_RB_CR} presents the collision rate caused by the lack of RBs in the target-SAT, resulting in \texttt{HO Request NACK}.
Our proposed DHO approach outperforms the two benchmark methods in terms of access delay and collision rate, especially when there are sufficient RBs available. 

To further investigate the impact of the number of RBs, in Table \ref{table:comparison_RB}, we present a comparison of the performance of our proposed DHO, under two scenarios: one in which enough RBs are available ($R_{k} = J, \ k \in \{1,2\}$)), and another in which the available RBs are insufficient ($R_{k} = 0.3 J, \ k \in \{1,2\}$)). 
While all methods eventually do a successful HO for a given RB, the proposed DHO achieves an average access delay that is 6.8x and 5.02x faster than conventional HO and random methods, respectively, when there are sufficient number of RBs available. Similarly, the DHO also enables ground UE to succeed in HO faster and with less collision when there is an insufficient number of RBs available.

Overall, the results of Fig. \ref{fig:comparion_RB} and Table \ref{table:comparison_RB} indicate that our proposed DHO exhibits a distinct advantage in terms of access delay, collision rate, and HO success rate when compared to the other two baselines. These findings emphasize the inherent flexibility of DHO in adapting and learning based on the specific network conditions, particularly with regard to the availability of RBs.

\subsubsection{Impact of Number of Preamble Signatures}
Fig. \ref{fig:comparion_PRACH} and Table \ref{table:comparison_PRACH} shows the impact of the number of RACH preamble signatures used in the RA process between the target-SAT and ground UEs, which is conducted after the UE receiving \texttt{HO Command}. The figure and table investigate the performance of HO methods in terms of access delay and collision rate.
The x-axis in Fig. \ref{fig:comparion_PRACH} represents the ratio of the number of preamble signatures to the number of ground UEs, indicating the proportion of preamble signatures available for the RA procedure. As the number of preamble signatures increases, the probability of PRACH contention (collision) decreases.

Still, the proposed DHO technique demonstrates superior performance in terms of access delay, achieving up to 4.83x and 2.59x faster than conventional HO and random methods, respectively. Interestingly, it, however, lags behind the random method in terms of PRACH collision rate, as shown in Fig. \ref{fig:comparion_PRACH_CR}. This can be attributed to the fact that DHO prioritizes access delay over collision rate, thus somewhat accepting a higher collision rate in order to achieve better access delay; this is more clearly indicated in Table \ref{table:comparison_PRACH}.

Overall, the results Figs. \ref{fig:comparion_RB} and \ref{fig:comparion_PRACH}, as well as Tables \ref{table:comparison_RB} and \ref{table:comparison_PRACH} provide validation for the effectiveness of our proposed DHO in selecting efficient actions based on specific network conditions, particularly in relation to the number of preamble signatures available. This DHO protocol flexibly behaves when resources are sufficient and when they are insufficient, as elaborated in the following subsection.


\subsection{DHO Protocol Behavior}
In Table \ref{table:agentPolicy}, the behavior of the DHO protocol is demonstrated. The results demonstrate that when resources are sufficient, DHO sends \texttt{HO Requests} to the greatest extent possible at every opportunity, while when resources are insufficient, DHO frequently opts not to send \texttt{HO Requests} for a few UEs.
This adaptive behavior of the DHO agent showcases its capability to adjust its actions based on network conditions, ultimately leading to optimal HO performance.

The behavior of the DHO agent is determined by the current network conditions, and it continuously updates its behavior as the conditions change over time through additional training\footnote{The use of \emph{transfer learning} techniques will allow our DHO agent, trained under specific network conditions, to adapt to new network conditions with only a few additional training episodes; this will be the focus of our future research.}. This ability to adapt to changing network conditions sets the DHO protocol apart from traditional HO methods, which may have limitations in adapting effectively to various network scenarios.

\subsection{Trade-off: Access Delay and Collision Rate} \label{sec:tradeoff}
The proposed DHO uses a reward function that considers both access delay and collision rate to train the serving-SAT's policy in selecting actions that minimize both factors.
However, these two factors are interrelated and present a trade-off. For instance, in order to minimize the collision rate, the number of \texttt{HO Request} attempts needs to be minimized, but this can increase the access delay. 

The importance placed on either access delay or collision rate may vary depending on the specific application. For instance, in applications requiring ultra-reliable low-latency communications (URLLC) capability, such as real-time wireless control and monitoring systems, minimizing access delay is crucial to ensure timely and reliable transmission of critical data. On the other hand, in massive machine-type communication (MMTC) scenarios, where a large number of devices are connected, minimizing collision rate becomes more important to maximize the overall network efficiency and capacity.

The proposed DHO enables the adjustment of the relative importance of these factors according to the specific requirements of the application, by changing the coefficients of the access delay and collision rate functions in the reward function, represented by $\nu$ in Table \ref{table:coeffRatio_Reward}\footnote{
The coefficient in the reward function $\nu$ is experimentally tuned to balance access delay and collision rate for optimal performance of the HO protocol.
}.

\subsection{Training and Convergence}


\begin{figure}[!h]
\vspace{-1.em}
\centering
\includegraphics[width=.95\columnwidth]{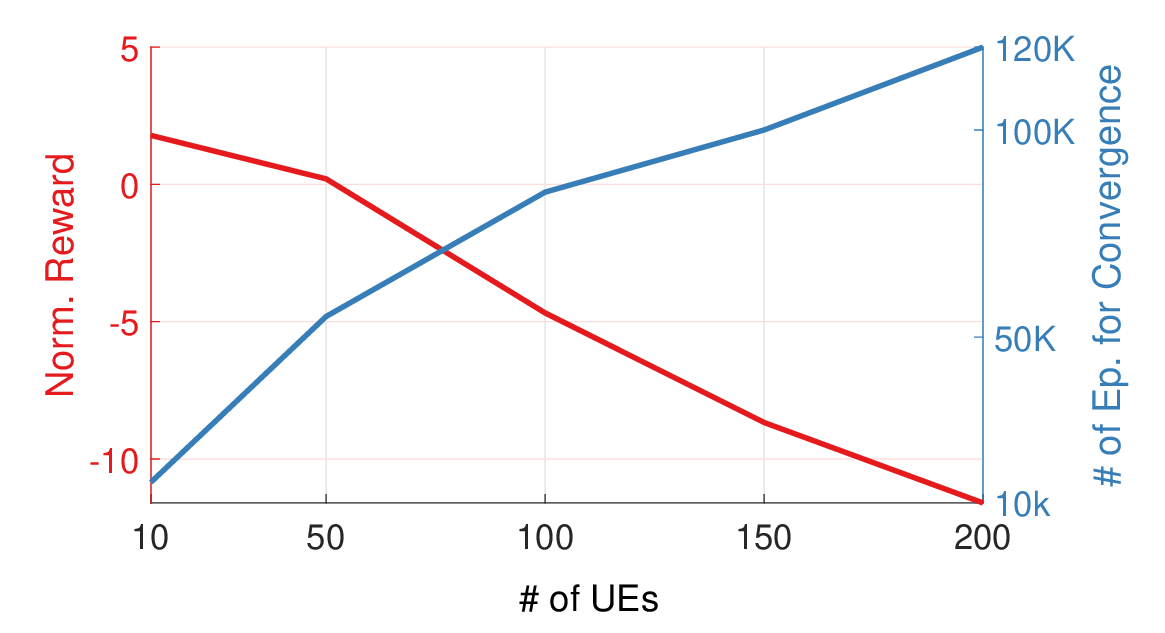}
\vspace{-1.em}
\caption{Impacts of the number of UE ($J=10$, $R_k = 0.5 J, \ \forall k$, and $P=2 J$).}
\label{fig:numberOfUE}
\end{figure}

\tpurp{
The training process for DRL can become increasingly difficult as the dimensions of the state and action spaces increase. This is especially relevant in our network scenario, where the number of UEs is a significant factor, making DRL learning more challenging. Specifically, the action space and state space of DHO are represented by $\mathbb{Z}^{J \times K}$ and $\mathbb{R}^{J \times (2J+1)}$, respectively. 
In general, as the dimensions of the state space and action space increase, the number of training steps required for the DRL agent to converge also increases. Furthermore, the learning reward tends to decrease with increasing dimensionality. 
Fig. \ref{fig:numberOfUE} shows the correlation between the number of UEs and the training step (episode) required. As shown in this result, training with $100$ UEs takes approximately five times longer than training with $10$ UEs. Our employed IMPALA algorithm is empirically known to be robust and stable in such a high-dimensional state and action space \cite{espeholt2018impala}. Our DRL-based HO algorithm can be efficiently scaled up to more UE cases using transfer learning or distributed training, which is deferred for future work.
}

%% file: Table_Result/Table_Parameter.tex
\begin{tabular} {l l}

\toprule[1pt]

\textbf{Environment Parameter} \cite{Starlink} & \\
\cmidrule(lr){1-1} \cmidrule(lr){2-2}

Region (area) of interest  & $1000 \times 1000$ {[}m$^2${]} \\
Number of UE & $J = 10 - 200$ \\
\# of orbital plane  & $3$ \\ 
Altitude of LEO SAT & $H_{\mathrm{L}}=550$ {[}km{]}  \\ 
Speed of LEO SAT & $ | \vb*{v} | = 7.59 \times 10^3$ {[}m/s{]} \\
Velocity of serving-SAT ($k=1$) & $\vb*{v}^{k} \in \big\{ [0, |\vb*{v}|, 0]^T \big\}$  \\
Velocity of target-SATs ($k=2,3$) & $\vb*{v}^{k} \in \Big\{ \Big[\frac{|\vb*{v}|}{\sqrt{2}}, \frac{|\vb*{v}|}{\sqrt{2}}, 0 \Big]^T, \Big[-\frac{|\vb*{v}|}{\sqrt{2}}, \frac{|\vb*{v}|}{\sqrt{2}}, 0 \Big]^T   \Big\}$  \\


\midrule
\textbf{HO Parameter} \cite{Handover_3GPP_LTE_2, Handover_3GPP_LTE_Parameter, Handover_3GPP_LTE_Parameter_2} & \\
\cmidrule(lr){1-1} \cmidrule(lr){2-2}

Period of measurement & $T_{\mathrm{M}} = 150$ [ms] \\
IIR filter order  & $k_{\mathrm{IIR}} = 4$ \\
Forgetting factor  & $\beta_{\mathrm{L_3}} = 0.5$  \\  
Period of HO update  & $T_{\mathrm{U}} = 300$ [ms] \\ 
Offset for A3 event & $1$ [dB] \\                                          



\midrule
\textbf{DRL Training Parameter} & \\
\cmidrule(lr){1-1} \cmidrule(lr){2-2}
Learning rate & $10^{-4}\sim5\times10^{-4}$ \\ 
Discount factor & $0.85\sim0.99$  \\ 
Batch size & $10000$ \\ 
\# of iterations per update & $10000$ \\ 
\# of iterations per episode & $20$ \\ 
\# of of episodes for training & $1000\sim8000$ \\
\bottomrule[1pt]
\end{tabular}

%% file: Table_Result/Table_RB.tex
\begin{tabularx}{1\linewidth}{l c c c}

\toprule[1pt]

\multicolumn{4}{c}{\textbf{Case \circled{1}: Enough RBs ($R_{k} = J, \ k \in \{1,2\}$)}} \\
\midrule[.5pt]
\textit{Schemes} & \textit{Avg. Access Delay}  & \textit{Avg. Collision (Insufficient RB)} & \textit{Avg. HO Success}  \\
\cmidrule(lr){1-1} \cmidrule(lr){2-2} \cmidrule(lr){3-3} \cmidrule(lr){4-4} 

HO (Conventional)
& $0.789$\; \tikz{
\draw[gray,line width=.3pt] (0,0) -- (1.1,0);
\draw[white, line width=0.01pt] (0,-2pt) -- (0,2pt);
\draw[black,line width=1pt] (0.925,0) -- (1.075,0);
\draw[black,line width=1pt] (0.925,-2pt) -- (0.925,2pt);
\draw[black,line width=1pt] (1.075,-2pt) -- (1.075,2pt);}
& $0$\; \tikz{
\draw[gray,line width=.3pt] (0,0) -- (1.1,0);
\draw[white, line width=0.01pt] (0,-2pt) -- (0,2pt);
\draw[black,line width=1pt] (0.0,-2pt) -- (0.0,2pt);
}
& $1$\; \tikz{
\draw[gray,line width=.3pt] (0,0) -- (1.1,0);
\draw[white, line width=0.01pt] (0,-2pt) -- (0,2pt);
\draw[black,line width=1pt] (1.075,-2pt) -- (1.075,2pt);}
\\

Random (Heuristic)
& $0.582$\; \tikz{
\draw[gray,line width=.3pt] (0,0) -- (1.1,0);
\draw[white, line width=0.01pt] (0,-2pt) -- (0,2pt);
\draw[black,line width=1pt] (0.535,0) -- (0.685,0);
\draw[black,line width=1pt] (0.535,-2pt) -- (0.535,2pt);
\draw[black,line width=1pt] (0.685,-2pt) -- (0.685,2pt);}
& $0$\; \tikz{
\draw[gray,line width=.3pt] (0,0) -- (1.1,0);
\draw[white, line width=0.01pt] (0,-2pt) -- (0,2pt);
\draw[black,line width=1pt] (0.0,-2pt) -- (0.0,2pt);
}
& $1$\; \tikz{
\draw[gray,line width=.3pt] (0,0) -- (1.1,0);
\draw[white, line width=0.01pt] (0,-2pt) -- (0,2pt);
\draw[black,line width=1pt] (1.075,-2pt) -- (1.075,2pt);}
\\

\textbf{DHO (Proposed)}
& $0.116$\; \tikz{
\draw[gray,line width=.3pt] (0,0) -- (1.1,0);
\draw[white, line width=0.01pt] (0,-2pt) -- (0,2pt);
\draw[black,line width=1pt] (0.071,0) -- (0.221,0);
\draw[black,line width=1pt] (0.071,-2pt) -- (0.071,2pt);
\draw[black,line width=1pt] (0.221,-2pt) -- (0.221,2pt);}
& $0$\; \tikz{
\draw[gray,line width=.3pt] (0,0) -- (1.1,0);
\draw[white, line width=0.01pt] (0,-2pt) -- (0,2pt);
\draw[black,line width=1pt] (0.0,-2pt) -- (0.0,2pt);
}
& $1$\; \tikz{
\draw[gray,line width=.3pt] (0,0) -- (1.1,0);
\draw[white, line width=0.01pt] (0,-2pt) -- (0,2pt);
\draw[black,line width=1pt] (1.075,-2pt) -- (1.075,2pt);}
\\

\midrule[0.8pt]
\multicolumn{4}{c}{\textbf{Case \circled{2}: Insufficient RBs ($R_{k} = 0.3 J, \ k \in \{1,2\}$)}} \\
\midrule[.5pt]
\textit{Schemes} & \textit{Avg. Access Delay}  & \textit{Avg. Collision (Insufficient RB)} & \textit{Avg. HO Success}  \\
\cmidrule(lr){1-1} \cmidrule(lr){2-2} \cmidrule(lr){3-3} \cmidrule(lr){4-4} 

HO (Conventiona)
& $13.52$\; \tikz{
\draw[gray,line width=.3pt] (0,0) -- (1.1,0);
\draw[white, line width=0.01pt] (0,-2pt) -- (0,2pt);
\draw[black,line width=1pt] (0.925,0) -- (1.075,0);
\draw[black,line width=1pt] (0.925,-2pt) -- (0.925,2pt);
\draw[black,line width=1pt] (1.075,-2pt) -- (1.075,2pt);}
& $12.7$\; \tikz{
\draw[gray,line width=.3pt] (0,0) -- (1.1,0);
\draw[white, line width=0.01pt] (0,-2pt) -- (0,2pt);
\draw[black,line width=1pt] (0.925,0) -- (1.075,0);
\draw[black,line width=1pt] (0.925,-2pt) -- (0.925,2pt);
\draw[black,line width=1pt] (1.075,-2pt) -- (1.075,2pt);}
& $0.59$\; \tikz{
\draw[gray,line width=.3pt] (0,0) -- (1.1,0);
\draw[white, line width=0.01pt] (0,-2pt) -- (0,2pt);
\draw[black,line width=1pt] (0.64,0) -- (0.73,0);
\draw[black,line width=1pt] (0.64,-2pt) -- (0.64,2pt);
\draw[black,line width=1pt] (0.73,-2pt) -- (0.73,2pt);}
\\

Random (Heuristic)
& $8.156$\; \tikz{
\draw[gray,line width=.3pt] (0,0) -- (1.1,0);
\draw[white, line width=0.01pt] (0,-2pt) -- (0,2pt);
\draw[black,line width=1pt] (0.428,0) -- (0.578,0);
\draw[black,line width=1pt] (0.428,-2pt) -- (0.428,2pt);
\draw[black,line width=1pt] (0.578,-2pt) -- (0.578,2pt);}
& $5.22$\; \tikz{
\draw[gray,line width=.3pt] (0,0) -- (1.1,0);
\draw[white, line width=0.01pt] (0,-2pt) -- (0,2pt);
\draw[black,line width=1pt] (0.324,0) -- (0.474,0);
\draw[black,line width=1pt] (0.324,-2pt) -- (0.324,2pt);
\draw[black,line width=1pt] (0.474,-2pt) -- (0.474,2pt);}
& $0.60$\; \tikz{
\draw[gray,line width=.3pt] (0,0) -- (1.1,0);
\draw[white, line width=0.01pt] (0,-2pt) -- (0,2pt);
\draw[black,line width=1pt] (0.64,0) -- (0.75,0);
\draw[black,line width=1pt] (0.64,-2pt) -- (0.64,2pt);
\draw[black,line width=1pt] (0.75,-2pt) -- (0.75,2pt);}
\\

\textbf{DHO (Proposed)}
& $8.053$\; \tikz{
\draw[gray,line width=.3pt] (0,0) -- (1.1,0);
\draw[white, line width=0.01pt] (0,-2pt) -- (0,2pt);
\draw[black,line width=1pt] (0.244,0) -- (0.394,0);
\draw[black,line width=1pt] (0.244,-2pt) -- (0.244,2pt);
\draw[black,line width=1pt] (0.394,-2pt) -- (0.394,2pt);}
& $4.15$\; \tikz{
\draw[gray,line width=.3pt] (0,0) -- (1.1,0);
\draw[white, line width=0.01pt] (0,-2pt) -- (0,2pt);
\draw[black,line width=1pt] (0.091,0) -- (0.241,0);
\draw[black,line width=1pt] (0.091,-2pt) -- (0.091,2pt);
\draw[black,line width=1pt] (0.241,-2pt) -- (0.241,2pt);}
& $0.60$\; \tikz{
\draw[gray,line width=.3pt] (0,0) -- (1.1,0);
\draw[white, line width=0.01pt] (0,-2pt) -- (0,2pt);
\draw[black,line width=1pt] (0.65,0) -- (0.75,0);
\draw[black,line width=1pt] (0.65,-2pt) -- (0.65,2pt);
\draw[black,line width=1pt] (0.75,-2pt) -- (0.75,2pt);}
\\

\bottomrule[1pt]
\end{tabularx}

%% file: Table_Result/Table_PRACH.tex
\begin{tabularx}{1\linewidth}{l c c }

\toprule[1pt]

\multicolumn{3}{c}{\textbf{Case \circled{3}: Enough preamble signatures ($P = 2 J$)}} \\
\midrule[.5pt]
\textit{Schemes} & \textit{Avg. Access Delay}  & \textit{Avg. Collision (PRACH Collision)}  \\
\cmidrule(lr){1-1} \cmidrule(lr){2-2} \cmidrule(lr){3-3} 

HO (Conventional)
& $1.231$\; \tikz{
\draw[gray,line width=.3pt] (0,0) -- (1.1,0);
\draw[white, line width=0.01pt] (0,-2pt) -- (0,2pt);
\draw[black,line width=1pt] (0.925,0) -- (1.075,0);
\draw[black,line width=1pt] (0.925,-2pt) -- (0.925,2pt);
\draw[black,line width=1pt] (1.075,-2pt) -- (1.075,2pt);}
& $3.31$\; \tikz{
\draw[gray,line width=.3pt] (0,0) -- (1.1,0);
\draw[white, line width=0.01pt] (0,-2pt) -- (0,2pt);
\draw[black,line width=1pt] (0.925,0) -- (1.075,0);
\draw[black,line width=1pt] (0.925,-2pt) -- (0.925,2pt);
\draw[black,line width=1pt] (1.075,-2pt) -- (1.075,2pt);}
\\

Random (Heuristic)
& $0.661$\; \tikz{
\draw[gray,line width=.3pt] (0,0) -- (1.1,0);
\draw[white, line width=0.01pt] (0,-2pt) -- (0,2pt);
\draw[black,line width=1pt] (0.462,0) -- (0.612,0);
\draw[black,line width=1pt] (0.462,-2pt) -- (0.462,2pt);
\draw[black,line width=1pt] (0.612,-2pt) -- (0.612,2pt);}
& $1.02$\; \tikz{
\draw[gray,line width=.3pt] (0,0) -- (1.1,0);
\draw[white, line width=0.01pt] (0,-2pt) -- (0,2pt);
\draw[black,line width=1pt] (0.227,0) -- (0.377,0);
\draw[black,line width=1pt] (0.227,-2pt) -- (0.227,2pt);
\draw[black,line width=1pt] (0.377,-2pt) -- (0.377,2pt);}
\\

\textbf{DHO (Proposed)}
& $0.255$\; \tikz{
\draw[gray,line width=.3pt] (0,0) -- (1.1,0);
\draw[white, line width=0.01pt] (0,-2pt) -- (0,2pt);
\draw[black,line width=1pt] (0.132,0) -- (0.282,0);
\draw[black,line width=1pt] (0.132,-2pt) -- (0.132,2pt);
\draw[black,line width=1pt] (0.282,-2pt) -- (0.282,2pt);}
& $2.13$\; \tikz{
\draw[gray,line width=.3pt] (0,0) -- (1.1,0);
\draw[white, line width=0.01pt] (0,-2pt) -- (0,2pt);
\draw[black,line width=1pt] (0.569,0) -- (0.719,0);
\draw[black,line width=1pt] (0.569,-2pt) -- (0.569,2pt);
\draw[black,line width=1pt] (0.719,-2pt) -- (0.719,2pt);}
\\

\midrule[0.8pt]
\multicolumn{3}{c}{\textbf{Case \circled{4}: Insufficient preamble signatures ($P = 0.8 J$)}} \\
\midrule[.5pt]
\textit{Schemes} & \textit{Avg. Access Delay}  & \textit{Avg. Collision (PRACH Collision)} \\
\cmidrule(lr){1-1} \cmidrule(lr){2-2} \cmidrule(lr){3-3} 

HO (Conventional)
& $1.740$\; \tikz{
\draw[gray,line width=.3pt] (0,0) -- (1.1,0);
\draw[white, line width=0.01pt] (0,-2pt) -- (0,2pt);
\draw[black,line width=1pt] (0.925,0) -- (1.075,0);
\draw[black,line width=1pt] (0.925,-2pt) -- (0.925,2pt);
\draw[black,line width=1pt] (1.075,-2pt) -- (1.075,2pt);}
& $9.41$\; \tikz{
\draw[gray,line width=.3pt] (0,0) -- (1.1,0);
\draw[white, line width=0.01pt] (0,-2pt) -- (0,2pt);
\draw[black,line width=1pt] (0.925,0) -- (1.075,0);
\draw[black,line width=1pt] (0.925,-2pt) -- (0.925,2pt);
\draw[black,line width=1pt] (1.075,-2pt) -- (1.075,2pt);}
\\

Random (Heuristic)
& $0.858$\; \tikz{
\draw[gray,line width=.3pt] (0,0) -- (1.1,0);
\draw[white, line width=0.01pt] (0,-2pt) -- (0,2pt);
\draw[black,line width=1pt] (0.418,0) -- (0.568,0);
\draw[black,line width=1pt] (0.418,-2pt) -- (0.418,2pt);
\draw[black,line width=1pt] (0.568,-2pt) -- (0.568,2pt);}
& $3.65$\; \tikz{
\draw[gray,line width=.3pt] (0,0) -- (1.1,0);
\draw[white, line width=0.01pt] (0,-2pt) -- (0,2pt);
\draw[black,line width=1pt] (0.313,0) -- (0.463,0);
\draw[black,line width=1pt] (0.313,-2pt) -- (0.313,2pt);
\draw[black,line width=1pt] (0.463,-2pt) -- (0.463,2pt);}
\\

\textbf{DHO (Proposed)}
& $0.648$\; \tikz{
\draw[gray,line width=.3pt] (0,0) -- (1.1,0);
\draw[white, line width=0.01pt] (0,-2pt) -- (0,2pt);
\draw[black,line width=1pt] (0.297,0) -- (0.447,0);
\draw[black,line width=1pt] (0.297,-2pt) -- (0.297,2pt);
\draw[black,line width=1pt] (0.447,-2pt) -- (0.447,2pt);}
& $4.24$\; \tikz{
\draw[gray,line width=.3pt] (0,0) -- (1.1,0);
\draw[white, line width=0.01pt] (0,-2pt) -- (0,2pt);
\draw[black,line width=1pt] (0.376,0) -- (0.526,0);
\draw[black,line width=1pt] (0.376,-2pt) -- (0.376,2pt);
\draw[black,line width=1pt] (0.526,-2pt) -- (0.526,2pt);}
\\

\bottomrule[1pt]
\end{tabularx}

%% file: Table_Result/Table_AgentPolicy.tex
\newcolumntype{R}{>{\raggedleft\arraybackslash}X}
\begin{tabularx}{1\linewidth}{l l l  }
    \toprule[1pt]
    \textit{Scenario} & \texttt{HO Request} for target-SATs ($a_j = 1, 2$) & No \texttt{HO Request} ($a_j = 0$) \\     
    \cmidrule(lr){1-1} \cmidrule(lr){2-2} \cmidrule(lr){3-3}
    Enough RB and PRACH (Case \circled{1} + \circled{3}) & \ \ \ \ \ \ \ \ \ \ \ \ {$93.7$ \%} \hspace{1pt}  \tikz{
        \fill[fill=color1] (0.0,0) rectangle (1.4055,0.2);
        \fill[pattern=north west lines, pattern color=black!30!color1] (0.0,0) rectangle (1.4055,0.2);
    } & \ \ \  {$6.33$ \%} \hspace{1pt}  \tikz{
        \fill[fill=color6] (0.0,0) rectangle (0.1,0.2);
        \fill[pattern=north west lines, pattern color=black!30!color6] (0.0,0) rectangle (0.1,0.2);
    } \\
    Insufficient RB and PRACH (Case \circled{2} + \circled{4}) & \ \ \ \ \ \ \ \ \ \ \ \  {$8.12$ \%} \hspace{1pt}  \tikz{
        \fill[fill=color2] (0.0,0) rectangle (0.2,0.2);
        \fill[pattern=north west lines, pattern color=black!30!color2] (0.0,0) rectangle (0.2,0.2);
    } & \ \ \  {$91.8$ \%} \hspace{1pt}  \tikz{
        \fill[fill=color7] (0.0,0) rectangle (1.377,0.2);
        \fill[pattern=north west lines, pattern color=black!30!color7] (0.0,0) rectangle (1.377,0.2);
    } \\
    \bottomrule[1pt]
\end{tabularx}


%% file: Table_Result/Table_Objective.tex
\begin{tabularx}{1\linewidth}{l c c }
\toprule[1pt]
{\textit{Objective}} & \textit{Avg. Access Delay} & \textit{Avg. Collision (PRACH Collision)}  \\
\cmidrule(lr){1-1} \cmidrule(lr){2-2} \cmidrule(lr){3-3} 

Delay-Aware ($\nu = 5$)  
& $0.0875$\; \tikz{
\draw[gray,line width=.3pt] (0,0) -- (1.1,0);
\draw[white, line width=0.01pt] (0,-2pt) -- (0,2pt);
\draw[black,line width=1pt] (0.178,0) -- (0.328,0);
\draw[black,line width=1pt] (0.178,-2pt) -- (0.178,2pt);
\draw[black,line width=1pt] (0.328,-2pt) -- (0.328,2pt);}
& $0.0845$\; \tikz{
\draw[gray,line width=.3pt] (0,0) -- (1.1,0);
\draw[white, line width=0.01pt] (0,-2pt) -- (0,2pt);
\draw[black,line width=1pt] (0.925,0) -- (1.075,0);
\draw[black,line width=1pt] (0.925,-2pt) -- (0.925,2pt);
\draw[black,line width=1pt] (1.075,-2pt) -- (1.075,2pt);}
\\

Collision-Aversion ($\nu = 1/20$) 
& $0.3461$\; \tikz{
\draw[gray,line width=.3pt] (0,0) -- (1.1,0);
\draw[white, line width=0.01pt] (0,-2pt) -- (0,2pt);
\draw[black,line width=1pt] (0.925,0) -- (1.075,0);
\draw[black,line width=1pt] (0.925,-2pt) -- (0.925,2pt);
\draw[black,line width=1pt] (1.075,-2pt) -- (1.075,2pt);}
& $0.0613$\; \tikz{
\draw[gray,line width=.3pt] (0,0) -- (1.1,0);
\draw[white, line width=0.01pt] (0,-2pt) -- (0,2pt);
\draw[black,line width=1pt] (0.65,0) -- (0.8,0);
\draw[black,line width=1pt] (0.65,-2pt) -- (0.65,2pt);
\draw[black,line width=1pt] (0.8,-2pt) -- (0.8,2pt);}
\\

\bottomrule[1pt]
\end{tabularx}

%% file: Section/Conclusion.tex
In conclusion, this work presents a novel HO protocol called DHO to address the challenge of massive access in LEO SAT networks. By using a DRL approach, the DHO protocol is able to minimize access delays and collision rates while simplifying the HO process. The numerical results demonstrate the superiority of the DHO protocol compared to conventional HO methods, with up to $6.8$x and $5.02$x lower access delay than conventional and heuristic methods, respectively.

As the next step, to fully realize the potential of regenerative-type LEO SAT networks, this will require optimization or a new design of existing specific functions in gNB in terrestrial networks, taking into account the unique characteristics of the NTN, such as their high-dynamic characteristics and massive access.